\documentclass[11pt]{article}

\usepackage[utf8]{inputenc}
\usepackage[T1]{fontenc}
\usepackage{lmodern}
\usepackage{graphicx}
\usepackage{amsmath}
\usepackage{amssymb}
\usepackage{booktabs}
\usepackage{hyperref}
\usepackage{multirow}
\usepackage{float}
\usepackage{enumitem}
\usepackage{url}
\usepackage{color}
\usepackage{listings}

\title{
Detecting Synthetic Political Narratives in Cross-Platform Social Media Discourse
}

\author{
Despoina Antonakaki$^{1,2}$ \and Sotiris Ioannidis$^{2}$
}

\newcommand{\affils}{%
\begin{center}
\small
$^{1}$Institute of Computer Science, Foundation for Research and Technology,\\
Vassilika Vouton, Heraklion, Crete, Greece\\[4pt]
$^{2}$Technical University of Crete, University Campus,\\
Akrotiri, Chania, Greece
\end{center}
}

\begin{document}

\maketitle
\affils

\begin{abstract}

The proliferation of large language models has introduced a new paradigm of synthetic political communication in which narratives may be generated, semantically coordinated, and strategically disseminated across platforms at scale. We present a cross-platform framework for detecting synthetic political narratives using four coordination signals --- lexical diversity D(C), temporal burstiness B(C), rhetorical repetition R(C), and semantic homogenization H(C) --- combined into a Synthetic Narrative Coordination Score SNC(C).

We apply the framework to a corpus of 353,223 records spanning six geopolitical event windows collected from six Telegram channels and nine Reddit communities (2023--2026). Results show that \textit{IntelSlava} exhibits the lowest lexical diversity (MATTR 0.52--0.54), the highest burstiness ($B=+0.48$ to $+0.73$), and the highest rhetorical overlap with peer channels (Jaccard 0.12), ranking first in the composite SNC(C) on four of six event windows (SNC 0.45--0.60). \textit{Rybar} ranks last on all windows despite its high semantic homogenization, because its Russian-language output yields high lexical diversity and near-zero rhetorical Jaccard with English-language channels --- demonstrating that no single indicator is sufficient for coordination detection. Multi-dimensional SNC(C) scoring provides a more robust and interpretable signal than any individual metric.

\end{abstract}


\section{Introduction}

Social media platforms have become central infrastructures for political communication, public opinion formation, and information dissemination during geopolitical crises and major world events. Platforms such as Telegram, Reddit, and X/Twitter enable the rapid propagation of narratives across geographically distributed communities, while simultaneously facilitating coordinated information campaigns and influence operations.
Traditional studies on online political discourse have primarily focused on sentiment analysis, topic modeling, misinformation detection, and behavioral bot identification. However, the emergence of large language models (LLMs) and generative AI systems has introduced a new paradigm of synthetic political communication in which narratives may be automatically generated, semantically coordinated, and strategically disseminated across platforms.

Recent advances in generative AI enable the creation of highly coherent and persuasive political content at unprecedented scale. Consequently, influence campaigns may increasingly rely on AI-assisted narrative generation, stylistic homogenization, and coordinated semantic amplification rather than solely on conventional automated bot behavior. These developments challenge existing detection methodologies that rely primarily on account-level behavioral indicators.

In this work, we investigate whether synthetic political discourse exhibits measurable collective characteristics that distinguish it from organic online discussions. Rather than attempting direct attribution of AI-generated text, we focus on identifying coordination patterns and synthetic dissemination signals across platforms.

The main contributions of this paper are summarized as follows:

\begin{itemize}
    \item We introduce a cross-platform framework for detecting synthetic political narrative coordination using four complementary signals: lexical diversity D(C), temporal burstiness B(C), rhetorical repetition R(C), and semantic homogenization H(C), combined into a composite SNC(C) score.

    \item We construct and release a corpus of 353,223 preprocessed records spanning six geopolitical event windows (2023--2026) collected from six Telegram channels and nine Reddit communities, including full Reddit comment trees.

    \item We demonstrate that no single coordination indicator is sufficient: a source (\textit{rybar}) that scores highest on semantic homogenization ranks last on the composite score, because high lexical diversity and near-zero cross-channel rhetorical overlap suppress its SNC(C) on all event windows.

    \item We show that discrete diplomatic events (Trump--Ukraine diplomacy) produce sharper coordination spikes than prolonged conflict windows, with three Telegram channels exceeding SNC\,=\,0.55 simultaneously.

    \item We provide open code and data to support reproducible research on cross-platform narrative coordination detection.
\end{itemize}

The remainder of this paper is organized as follows. Section~2 reviews related work. Section~3 presents the dataset, preprocessing pipeline, and all coordination analyses with results. Section~4 describes the methodology formally. Section~5 presents the experimental setup. Section~6 summarises the key findings. Section~7 discusses implications. Section~8 acknowledges limitations. Section~9 concludes.

\section{Related Work}

\subsection{Political Discourse Analysis}

Computational political discourse analysis has traditionally examined how public opinion, ideology, and political framing emerge in large text collections. Social media studies commonly use sentiment analysis, stance detection, topic modeling, and network analysis to identify polarization, agenda setting, and the diffusion of political claims \cite{antonakaki2021survey,antonakaki2017political,antonakaki2016referendum}. During geopolitical crises, these methods are especially useful because the same event can generate competing interpretations across different communities, languages, and media ecosystems \cite{antonakaki2025israelhamaswar,antonakaki2025crossplatformisrael,antonakaki2026iran}.
However, event-centered political discourse analysis often treats platforms independently. A narrative that appears organically fragmented on one platform may look coordinated only when its cross-platform timing, wording, and semantic similarity are examined together. This motivates the cross-platform design of the present study \cite{antonakaki2026venezuela}.

\subsection{Bot Detection in Social Media}

Bot detection research focuses on identifying automated or semi-automated accounts using account metadata, posting frequency, reposting behavior, follower graphs, and temporal regularities \cite{shevtsov2022bots,shevtsov2023russiaukraine,lamprou2024crisis}. These approaches are valuable when manipulation is driven by automated accounts, but they are less reliable when influence operations use human-operated accounts, mixed human-machine workflows, or AI-assisted text generation. In such cases, suspicious behavior may appear at the collective narrative level rather than at the level of a single account. Recent work on propaganda and adversarial information operations similarly emphasizes coordinated behavior, cross-platform propagation, and multimodal manipulation as features of modern influence campaigns \cite{polychronis2025propaganda,silalahi2026multimodal}.

Our work therefore does not define synthetic political discourse as text produced by a specific model or bot. Instead, we operationalize synthetic dissemination as an observable pattern of coordination: unusually similar messages, repeated rhetorical templates, compressed timing, and aligned emotional framing across communities.

\subsection{AI-Generated Text Detection}

AI-generated text detection is commonly framed as a document-level classification task. Existing detectors often rely on token probability patterns, perplexity, burstiness, or supervised classifiers trained on model-generated and human-written samples. While useful in constrained settings, these methods can degrade across languages, domains, model families, prompt styles, and paraphrasing strategies. Political influence content is particularly difficult because it may be edited by humans, translated, summarized, or mixed with authentic quotations and links.

For this reason, the proposed framework avoids direct authorship attribution. We instead examine whether sets of messages exhibit collective properties associated with synthetic dissemination: semantic homogenization, rhetorical repetition, low linguistic diversity, and synchronized propagation around salient events.

\subsection{LLMs, Disinformation, and Propaganda Analysis}

Recent work increasingly studies LLMs as both analytic tools for detecting information disorder and potential generators of persuasive misinformation. Inman \cite{inman2025disinformation} is especially close to our semantic framing: the thesis proposes measuring continuous similarity between social media posts and target disinformation narratives, then using clustering to synthesize dominant narrative variants.

Maurya et al. \cite{maurya2025simulating} use LLM personas as synthetic agents to simulate misinformation diffusion through social networks, showing how identity- and ideology-conditioned rewriting can degrade factual fidelity across propagation chains. This motivates our interest in temporal and semantic drift, although our study measures real platform data rather than simulated diffusion.
Saxena \cite{saxena2024llms} frames LLMs as tools for curtailing disinformation, emphasizing their ability to analyze large text collections and support detection workflows. This aligns with our use of LLM-compatible semantic representations, but our paper focuses on cross-platform coordination signals rather than single-message classification.
Kuntur et al. \cite{kuntur2025under} survey the use of LLMs in fake news detection and summarize the field's shift from conventional supervised classifiers toward generative, explanation-oriented, and retrieval-augmented approaches. Their review supports our decision to position the paper within the LLM-era detection literature while avoiding brittle direct AI authorship claims.

Ahmad et al. \cite{ahmad2025enhanced} fine-tune deep learning models for propaganda detection in public social media discussions, emphasizing multi-label and multi-class identification of propaganda techniques. Their work is relevant to the rhetorical-repetition part of our framework, while our study treats propaganda-like language as one signal among semantic and temporal coordination features.
Barfar and Sommerfeldt \cite{barfar2026propaganda} analyze propaganda-style content generated by GPT models and trace hidden linguistic strategies across model versions. Their finding that LLM-generated propaganda can exhibit coherent rhetorical patterns motivates our linguistic-entropy and rhetorical-template features.
Dabiriaghdam \cite{dabiriaghdam2025combating} studies generative-AI disinformation defenses, including watermarking LLM outputs and multimodal persuasion analysis of memes. This work is adjacent to our project because it emphasizes robustness to paraphrasing and multimodal persuasion, both of which are important future extensions of the current text-first dataset.
Shah et al. \cite{shah2024doubleedged} describe LLMs as double-edged tools that can both generate disinformation and support detection and mitigation. Their framing supports the central premise of our paper: synthetic political narratives should be studied as an ecosystem-level risk rather than only as isolated generated texts.

Ahmad et al. \cite{ahmad2026irlm} propose an information-retrieval language-model framework for zero- and few-shot propaganda detection in social media content. This is relevant to our planned expansion because few-shot retrieval-style methods may help label narrative clusters without requiring a large manually annotated training set.
AI \cite{ai2025trustworthy} develops a broader trustworthy-AI framework for detecting, understanding, and mitigating information disorder across textual, visual, and propagation-based features. This dissertation provides a useful umbrella reference for our modular design, which also separates detection, interpretation, and mitigation-oriented analysis.
Kwon and Jang \cite{kwon2025comprehensive} survey fake-text detection across both misinformation and language-model-generated text, explicitly connecting misinformation detection with LM-generated text detection. Their taxonomy reinforces our choice to treat synthetic political narratives as a hybrid problem involving both misleading content and generated or paraphrased language.
Donner \cite{donner2024misinformation} evaluates misinformation detection methods using LLMs and application programming interfaces. This is relevant to our implementation planning because the proposed pipeline may later compare open-source embedding models with API-based LLM judgments for cluster validation.

Gambini \cite{gambini2024digital} studies social-media defense strategies against LLM-generated content and reports work on deepfake text detection for short social-media posts. This is directly relevant to our Telegram and Reddit setting, where messages are short, noisy, and often harder to classify than long-form articles.
Papageorgiou et al. \cite{papageorgiou2024survey} survey LLM use in fake news and fake profile detection, covering fact-checking, text classification, and social-media applications. Their review supports the broader motivation for using LLM-compatible features while also noting challenges of interpretability, data bias, and scalability.
Sallami et al. \cite{sallami2024deception,aimeur2025socially} frame LLMs as both generators and detectors of fake news and argue that socially responsible fake-news detection requires more than optimizing accuracy. This perspective is important for our paper because coordination detection may affect democratic discourse and therefore requires cautious interpretation and human validation.
Burmester \cite{burmester2026belief} uses LLM-based agent networks to simulate bot-driven belief manipulation in a Twitter-like environment. This complements Maurya et al.'s simulation work and motivates our interest in measuring whether real cross-platform narratives exhibit synchronized timing and repeated persuasive frames.
Bontcheva et al. \cite{bontcheva2024generative} provide a broad white paper on generative AI and disinformation, covering creation, spread, detection, and debunking. Their report strengthens our framing of synthetic narrative detection as an ecosystem problem involving generation, diffusion, and platform-level mitigation.

\subsection{Cross-Platform Narrative Analysis}

Cross-platform narrative analysis studies how stories, frames, hashtags, claims, and communities move between online spaces. Telegram is important for this paper because public channels often operate as broadcast infrastructures for political messaging and can preserve chronological streams of posts. Reddit is useful as a comparison platform because it contains threaded discussion, visible community boundaries, and a mix of highly moderated and loosely moderated political forums. X/Twitter remains relevant to political diffusion studies, but recent access constraints make it less suitable as the primary data source for a first reproducible dataset.

Prior work by the same authors has studied cross-platform political discourse on Telegram, Reddit, and X/Twitter during the Israel--Hamas conflict, the Iran--Israel escalation, and political turbulence in Venezuela, establishing coordinated information dissemination as a consistent cross-platform phenomenon \cite{antonakaki2025israelhamaswar,antonakaki2025crossplatformisrael,antonakaki2026iran,antonakaki2026venezuela}. Cinus et al. \cite{cinus2025crossplatform} provide a close methodological reference point: they analyze cross-platform coordinated inauthentic activity around the 2024 U.S. election using Facebook, X/Twitter, and Telegram data, constructing similarity networks around shared URLs and similar textual content. Bawa et al. \cite{bawa2025telegram} provide a closer domain reference for our Telegram component, releasing a large-scale Russia--Ukraine Telegram corpus spanning pro-Kremlin and anti-Kremlin channels. Our study differs by pairing Telegram broadcast channels with Reddit discussion communities and by focusing on semantic, rhetorical, and temporal signals associated with synthetic narrative dissemination rather than only platform behavior or URL co-sharing.

Accordingly, this paper uses Telegram and Reddit as the core dataset sources. X/Twitter is treated as an optional extension for future validation if compliant API access is available.

\section{Dataset Collection and Preprocessing}
\label{sec:dataset}

\subsection{Platforms and Data Sources}

The dataset is built on two complementary public data sources: Telegram public channels and Reddit public communities. Telegram provides broadcast-style political posts, forwarded messages, links, and channel-level timelines. Reddit provides submissions and full threaded comment trees organised by community, score, and timestamp. Together, these two platforms allow us to compare top-down narrative broadcasting with bottom-up participatory discussion across the same geopolitical event windows.

We exclude private groups, direct messages, deleted content, and content requiring circumvention of platform access controls. The collection stores only the fields required for narrative analysis: platform identifier, source name, message or post identifier, timestamp, text body, URL metadata, reply or thread identifier, engagement counters (views and forwards on Telegram; score and comment count on Reddit), and author pseudonyms. Author names are not used as analytical features; the unit of analysis is the message and the source channel or subreddit.

\begin{table}[H]
\centering
\resizebox{\linewidth}{!}{%
\begin{tabular}{p{0.15\linewidth}p{0.25\linewidth}p{0.32\linewidth}p{0.20\linewidth}}
\toprule
\textbf{Platform} & \textbf{Source} & \textbf{Editorial stance / focus} & \textbf{Subscribers} \\
\midrule
\multirow{6}{*}{Telegram}
 & \path{KyivIndependent_official} & Pro-Ukrainian English-language news          & $>$1.4 M \\
 & wartranslated             & Ukraine war frontline translation            & $>$200 K \\
 & rybar                     & Pro-Russian military analysis (RU)           & $>$1.3 M \\
 & IntelSlava                & Pro-Russian intelligence aggregator (RU/EN)  & $>$1.5 M \\
 & \path{MiddleEastEye_TG}   & Middle East news, critical of Western policy  & $>$500 K \\
 & DDGeopolitics             & Multi-conflict geopolitics aggregator         & $>$700 K \\
\midrule
\multirow{9}{*}{Reddit}
 & r/worldnews               & General international news discussion  & $>$32 M \\
 & r/geopolitics             & Academic-style geopolitics             & $>$2 M  \\
 & r/worldpolitics           & International politics commentary      & $>$1.1 M \\
 & r/news                    & General news                           & $>$30 M \\
 & r/ukraine                 & Ukraine conflict, pro-Ukrainian leaning & $>$2.3 M \\
 & r/europe                  & European affairs                       & $>$3 M  \\
 & r/credibledefense         & Military and defence analysis          & $>$580 K \\
 & r/IsraelPalestine         & Israel--Palestine debate               & $>$350 K \\
 & r/PoliticalDiscussion     & Structured political discussion        & $>$180 K \\
\bottomrule
\end{tabular}}
\caption{Data sources included in the corpus. Telegram subscriber counts and Reddit member counts are approximate figures at collection time. The six Telegram channels span both Western-aligned and pro-Russian editorial stances, enabling cross-narrative comparison.}
\label{tab:data_sources}
\end{table}

The source selection deliberately includes channels from opposing editorial positions. \textit{Rybar} and \textit{Intel Slava} are well-documented pro-Kremlin channels that publish primarily in Russian; \textit{KyivIndependent\_official} and \textit{wartranslated} represent the Ukrainian and Western-aligned perspective; \textit{MiddleEastEye} and \textit{DDGeopolitics} cover Middle Eastern and multipolar geopolitical framing. This ideological spread is intentional: synthetic coordination signals are more robust when they appear across sources with otherwise divergent editorial stances.

\subsection{Event Windows}

Data collection is organized around six geopolitical event windows rather than open-ended keyword scraping. This design constrains the corpus to politically coherent narrative clusters, makes temporal coordination measurable against known event timestamps, and enables pre/during/post comparisons where a reference date $t_0$ exists.

\begin{table}[H]
\centering
\resizebox{\linewidth}{!}{%
\begin{tabular}{lp{0.32\linewidth}ccc}
\toprule
\textbf{Event ID} & \textbf{Description} & \textbf{Start} & \textbf{End} & \textbf{$t_0$} \\
\midrule
ukraine\_war\_general    & Russia--Ukraine war discourse (full year)          & 2025-05-15 & 2026-05-14 & --- \\
israel\_gaza\_general    & Israel--Gaza conflict discourse (full year)        & 2025-05-15 & 2026-05-14 & --- \\
iran\_israel\_escalation & Iran--Israel direct military confrontations        & 2024-04-01 & 2026-05-14 & 2024-04-13 \\
gaza\_conflict\_full     & Gaza conflict from the Oct.~7 Hamas attack onwards & 2023-10-01 & 2026-05-14 & 2023-10-07 \\
trump\_ukraine\_diplomacy & Trump second-term US--Ukraine--Russia diplomacy   & 2025-01-20 & 2026-05-14 & --- \\
nato\_summit\_2025       & NATO Summit, The Hague, June 2025                  & 2025-06-11 & 2025-07-09 & 2025-06-25 \\
\bottomrule
\end{tabular}}
\caption{Event windows used for data collection. Events with a $t_0$ anchor support three-period analysis (pre-event, acute, post-event). Events without a $t_0$ are treated as continuous monitoring windows spanning the collection period.}
\label{tab:event_windows}
\end{table}

For events with a $t_0$ anchor, messages in the interval $[t_0, t_0+14\,\text{d}]$ are labelled the \textit{acute period}; messages before $t_0$ serve as a pre-event baseline; messages after $t_0+14\,\text{d}$ form the post-event diffusion window. This structure enables the temporal synchronization metric $T(C)$ to be calibrated against pre-event posting rhythms.

Each event has an associated multilingual keyword list (English and Russian) covering actor names, place names, hashtags, and event-specific phrases. The keyword list is applied as a post-hoc filter: all messages from the selected sources are retrieved for the collection window and then filtered to retain only those matching at least one keyword. This avoids the selection bias that would arise from keyword-restricted API queries while keeping the corpus thematically focused.

\subsection{Telegram Data Collection}

Telegram collection uses the Telegram API through the \texttt{Telethon} library, authenticating with a researcher-registered API credential. For each channel listed in Table~\ref{tab:data_sources}, the collector iterates over all messages in reverse chronological order using \texttt{iter\_messages}, stopping when the message timestamp falls below the event window start date. Each message is stored as a JSON record containing: \texttt{platform}, \texttt{event\_id}, \texttt{source} (channel username), \texttt{message\_id}, \texttt{timestamp} (UTC ISO-8601), \texttt{text}, \texttt{views}, \texttt{forwards}, and \texttt{reply\_to}.

The Telegram component is especially important for detecting broadcast coordination. Pro-Russian channels (\textit{Rybar}, \textit{Intel Slava}) publish in Russian and provide a bilingual signal: coordinated narratives may appear in Russian on Telegram before being translated and amplified in English on Reddit. We do not collect private chats, group chats, or any content requiring authentication beyond the researcher API credential.

\subsection{Reddit Data Collection}

Reddit collection targets the public subreddits in Table~\ref{tab:data_sources} using the Reddit public JSON search endpoint with keyword queries. For each event and subreddit, the collector submits paginated \texttt{/search.json} queries with \texttt{sort=new\&t=all}, retrieving up to 100 submissions per page with exponential back-off on rate-limit responses. Each submission is then expanded to include its full comment tree, retrieved recursively from the \texttt{/comments} endpoint. This yields two record types per post: a submission record (title, body text, score, URL, author, timestamp) and one comment record per reply (body text, score, parent ID, depth, timestamp).

The Reddit public search API does not reliably return results older than approximately six months for narrow queries, so historical depth is inherently limited for specific event windows. For the two large ongoing conflict windows (\textit{ukraine\_war\_general} and \textit{israel\_gaza\_general}), which use broad keyword sets, the API returned sufficient volume. The four narrower event windows (Iran--Israel escalation, Gaza conflict full timeline, Trump--Ukraine diplomacy, NATO Summit 2025) yielded no Reddit results and are analysed using Telegram data only.

Reddit serves a complementary role to Telegram in the analysis. While Telegram channels are curated broadcast sources with clear editorial stances, Reddit reflects community-level discussion, counter-narratives, and organic information diffusion. Narratives that appear first in Telegram broadcasts and later surface in Reddit threads with similar framing provide evidence for cross-platform propagation.

\subsection{Dataset Scale and Comparison}

\begin{table}[H]
\centering
\resizebox{\linewidth}{!}{%
\begin{tabular}{p{0.22\linewidth}p{0.16\linewidth}p{0.30\linewidth}p{0.24\linewidth}}
\toprule
\textbf{Study} & \textbf{Platforms} & \textbf{Scope} & \textbf{Scale} \\
\midrule
Cinus et al. \cite{cinus2025crossplatform} & Facebook, X, Telegram & 2024 U.S. election (May--June 2024) & 46K Facebook; 6M X; 4.3M Telegram \\
Bawa et al. \cite{bawa2025telegram} & Telegram & Russia--Ukraine conflict (2021--2023) & 518 channels; 5.2M posts \\
Polychronis \cite{polychronis2025propaganda} & Multiple & State-sponsored propaganda analysis & Not publicly available \\
Silalahi et al. \cite{silalahi2026multimodal} & Multiple & Multimodal influence operations & Not publicly available \\
\midrule
\textbf{This work} & Telegram, Reddit & Six geopolitical windows (2023--2026) & \textbf{50,777 Telegram; 321,387 Reddit; 372,164 total} \\
\bottomrule
\end{tabular}}
\caption{Comparison of related information-operation datasets with the present corpus. Our dataset is the only one combining Telegram broadcast channels with full Reddit comment trees across multiple geopolitical event windows.}
\label{tab:related_dataset_comparison}
\end{table}

Our corpus is smaller than large-scale single-platform collections such as Bawa et al.\ \cite{bawa2025telegram}, which covers 518 Telegram channels, but it is the first to combine Telegram broadcast content with full Reddit discussion trees across six geopolitical event windows spanning 2023--2026. The cross-platform design is the primary contribution: single-platform studies cannot observe whether narratives migrate between broadcast and discussion environments, which is central to the coordination detection task.

\subsection{Preprocessing Pipeline}

The preprocessing pipeline normalizes platform-specific records into a common schema. Each record carries a unified timestamp, a \texttt{\_text} field concatenating all available text columns (title, body, message text), a \texttt{char\_len} and \texttt{word\_len} field computed after stripping whitespace, a \texttt{platform} label, an \texttt{event\_id}, and a \texttt{source} identifier. Telegram records carry \texttt{views} and \texttt{forwards}; Reddit records carry \texttt{score} and \texttt{num\_comments}.

Text is lowercased only for lexical matching and deduplication; original casing is preserved for stylometric and embedding analysis. URLs, hashtags, and emojis are retained because they carry coordination signals. Boilerplate, empty records, media-only messages (zero-length text), duplicate API records, and messages below ten characters are removed.

Language identification is applied using a Cyrillic-character ratio heuristic: messages in which more than 15\% of characters are Cyrillic are labelled Russian; all others are labelled English or unknown. The corpus is intentionally bilingual for Telegram: Russian-language content from \textit{Rybar} and \textit{Intel Slava} is retained alongside English content because cross-lingual narrative alignment is a target signal for the coordination analysis. Near-duplicate detection is performed using normalised text hashes; exact duplicates and near-duplicates are flagged (not removed) so that repetition frequency itself can be used as a coordination feature. The final dataset contains raw normalised records and will be extended with derived feature tables for embeddings, temporal bins, and narrative clusters during the analysis phase.

\subsection{Dataset Volume Analysis}

Before proceeding to semantic analysis, we perform a volume analysis of the collected corpus to characterize its temporal coverage, source distribution, linguistic composition, and posting dynamics. This analysis serves three purposes: it validates the collection infrastructure, it provides baseline statistics for the paper's descriptive tables, and it yields early coordination signals visible at the aggregate level before any embedding computation.

\subsubsection{Message Volume and Temporal Coverage}

Table~\ref{tab:volume_summary} summarises the record counts, source coverage, and temporal span for the two primary event windows. Telegram provides denser, longer-span coverage because the \texttt{iter\_messages} API retrieves the full channel history back to the configured start date. Reddit coverage is shallower in historical depth due to API access constraints on the public search endpoint, but compensates in volume through the collection of full comment trees under each submission.

\begin{table}[H]
\centering
\resizebox{\linewidth}{!}{%
\begin{tabular}{lcrrrr}
\toprule
\textbf{Event} & \textbf{Platform} & \textbf{Records} & \textbf{Sources} & \textbf{Span (d)} & \textbf{Avg/day} \\
\midrule
ukraine\_war\_general    & Telegram & 32{,}237  & 6 & 364 &  88.6 \\
ukraine\_war\_general    & Reddit   & 166{,}999 & 9 & 219 & 762.6 \\
israel\_gaza\_general    & Telegram & 12{,}043  & 6 & 364 &  33.1 \\
israel\_gaza\_general    & Reddit   & 154{,}388 & 8 & 285 & 541.7 \\
\midrule
iran\_israel\_escalation & Telegram &  4{,}435  & 6 & 773 &   5.7 \\
gaza\_conflict\_full     & Telegram &  1{,}564  & 6 & 951 &   1.6 \\
trump\_ukraine\_diplomacy & Telegram &   369    & 6 & 476 &   0.8 \\
nato\_summit\_2025       & Telegram &   125    & 6 &  28 &   4.5 \\
\midrule
\textbf{Total}           & ---      & \textbf{372{,}140} & --- & --- & --- \\
\bottomrule
\end{tabular}}
\caption{Volume summary for all event windows. Reddit records include both submissions and full comment trees (only two events had Reddit data). Telegram records are individual channel messages. The Reddit search API returned results only for the two large ongoing conflict windows; specific event windows returned no Reddit data.}
\label{tab:volume_summary}
\end{table}

The higher average daily volume on Reddit reflects the hierarchical comment structure: each submission typically generates tens to hundreds of comments, all retained for linguistic analysis. On Telegram, the average of 88.6 messages per day for the Ukraine window is consistent with active political broadcast channels producing multiple posts per hour during peak news periods. The four smaller Telegram-only event windows (\textit{iran\_israel\_escalation}, \textit{gaza\_conflict\_full}, \textit{trump\_ukraine\_diplomacy}, \textit{nato\_summit\_2025}) yielded no Reddit results because the Reddit public search API does not reliably surface older content for narrow keyword queries; these windows are analysed using Telegram data only. The total corpus across all platforms and events comprises approximately 372{,}000 records.

\subsubsection{Language Composition}

A Cyrillic-character heuristic applied to message text reveals that Russian-language content constitutes approximately 13\% (4,074 messages) of the Telegram Ukraine corpus and 10\% (1,173 messages) of the Telegram Gaza corpus, contributed primarily by the \textit{Rybar} and \textit{Intel Slava} channels. For the \textit{iran\_israel\_escalation} window, Russian-language content rises to 34\% (1,509 of 4,435 messages), reflecting these channels' consistent coverage of regional flashpoints. Reddit content is almost exclusively English ($<$0.1\% Cyrillic across both windows). This bilingual structure is deliberate: comparing English-language and Russian-language narrative framing of the same events is central to the coordination detection framework, since coordinated synthetic campaigns may exhibit parallel narrative templates across language groups.

\subsubsection{Cumulative Distribution Analysis}

\begin{figure}[H]
\centering
\includegraphics[width=\linewidth]{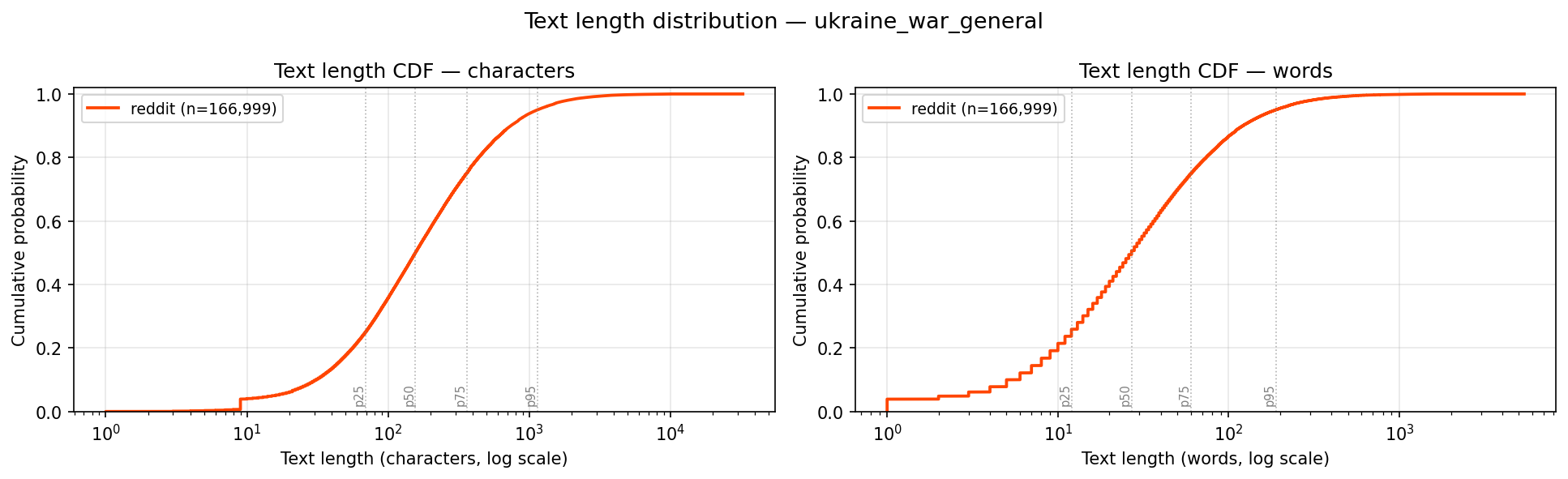}
\caption{Text length ECDF for the \textit{ukraine\_war\_general} event window. Left: character count; right: word count. Both axes are log-scaled. Vertical dotted lines mark the 25th, 50th, 75th, and 95th percentiles. Telegram messages are shorter and more uniform; Reddit spans a wider range due to comment-tree variation.}
\label{fig:cdf_textlen_ukraine}
\end{figure}

\begin{figure}[H]
\centering
\includegraphics[width=\linewidth]{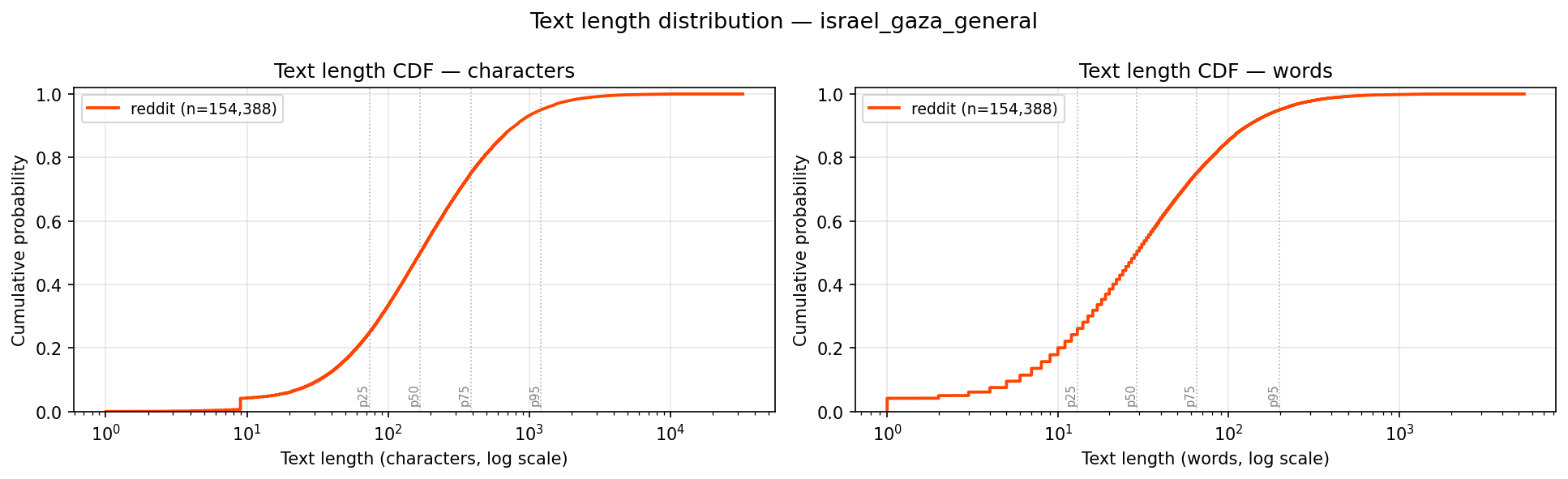}
\caption{Text length ECDF for the \textit{israel\_gaza\_general} event window. The pattern mirrors the Ukraine window, with Telegram showing a narrower, left-shifted distribution relative to Reddit.}
\label{fig:cdf_textlen_gaza}
\end{figure}

\begin{figure}[H]
\centering
\includegraphics[width=\linewidth]{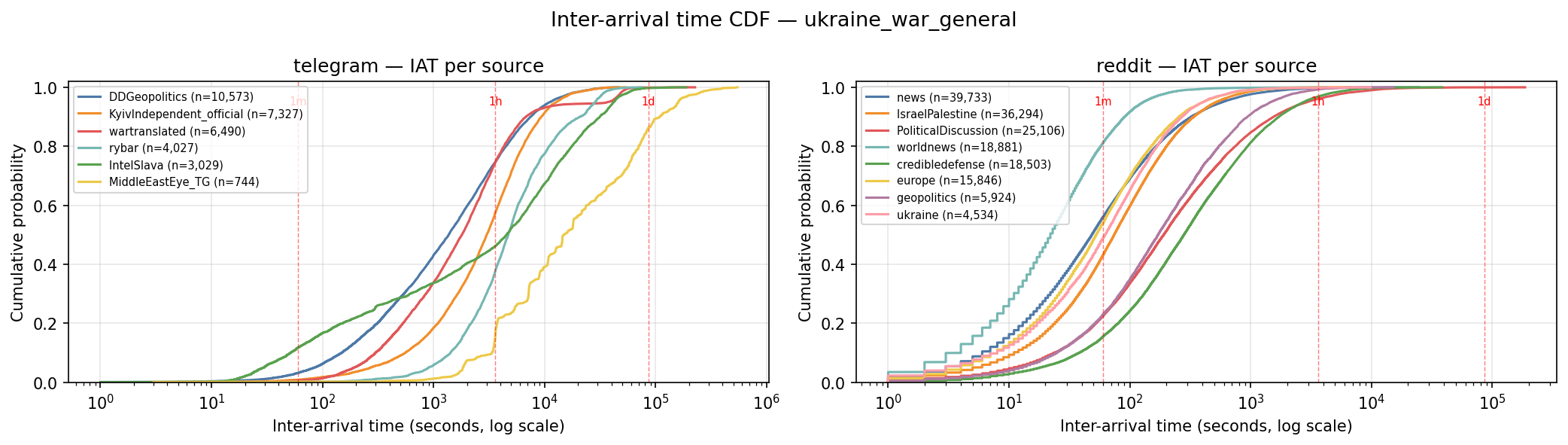}
\caption{Inter-arrival time (IAT) ECDF per source for the \textit{ukraine\_war\_general} window. Each curve represents one channel or subreddit. Red dashed lines mark 1-minute, 1-hour, and 1-day thresholds. Sources with a left-shifted curve post in rapid bursts; right-shifted sources post intermittently. Burst behaviour is a candidate coordination signal.}
\label{fig:cdf_iat_ukraine}
\end{figure}

\begin{figure}[H]
\centering
\includegraphics[width=\linewidth]{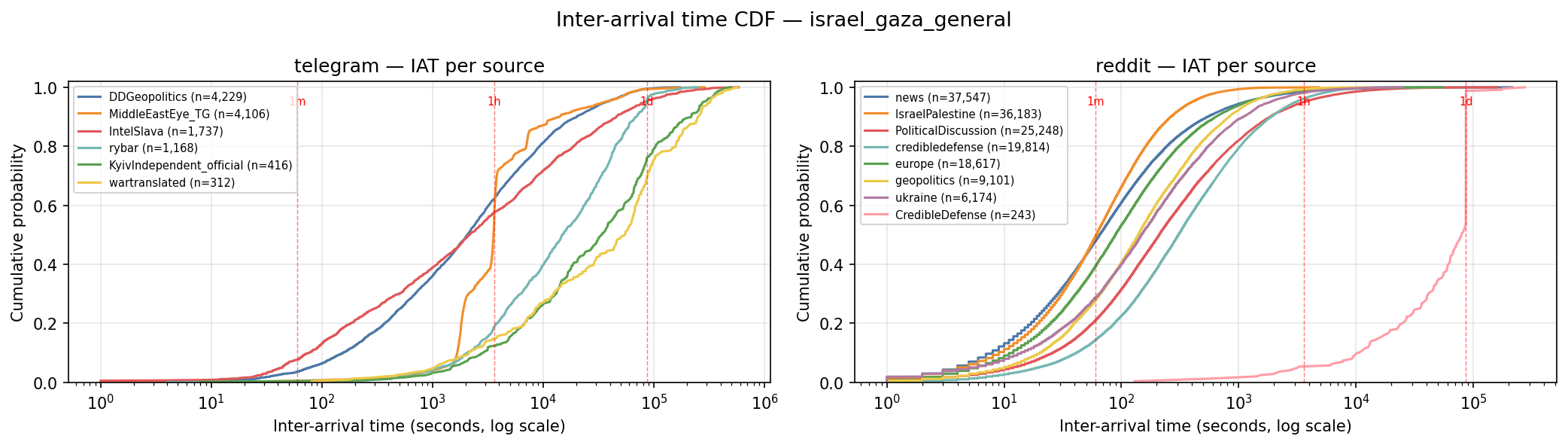}
\caption{Inter-arrival time ECDF for the \textit{israel\_gaza\_general} window. Middle East Eye and DDGeopolitics show the most active posting rhythms for this topic.}
\label{fig:cdf_iat_gaza}
\end{figure}

\begin{figure}[H]
\centering
\includegraphics[width=\linewidth]{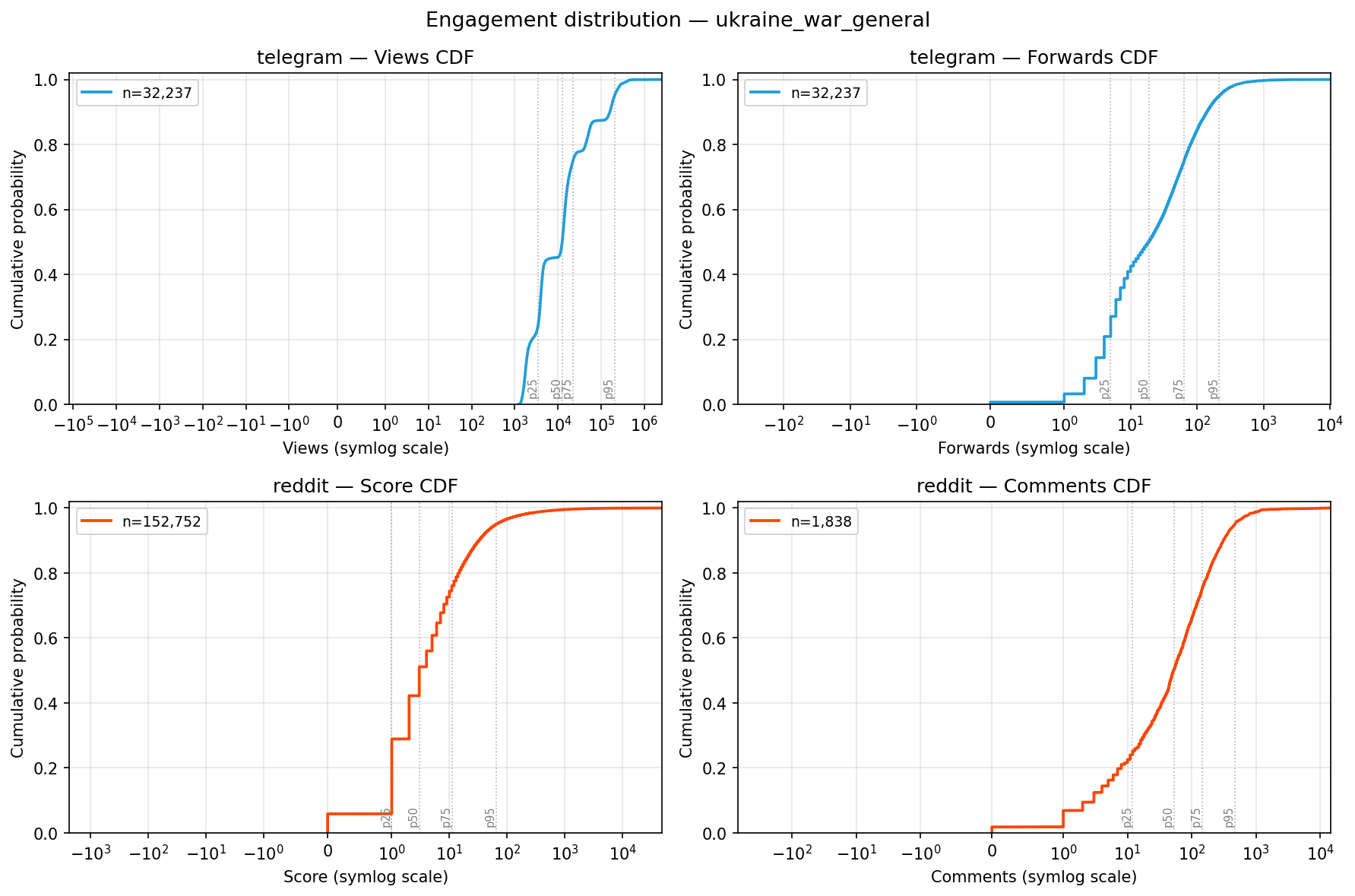}
\caption{Engagement distribution ECDFs for the \textit{ukraine\_war\_general} window. Top row: Telegram views and forwards. Bottom row: Reddit score and comment count. Both Telegram metrics follow heavy-tailed distributions; Reddit engagement is more zero-inflated. Axes use symlog scaling to accommodate zero values.}
\label{fig:cdf_engagement_ukraine}
\end{figure}

\begin{figure}[H]
\centering
\includegraphics[width=\linewidth]{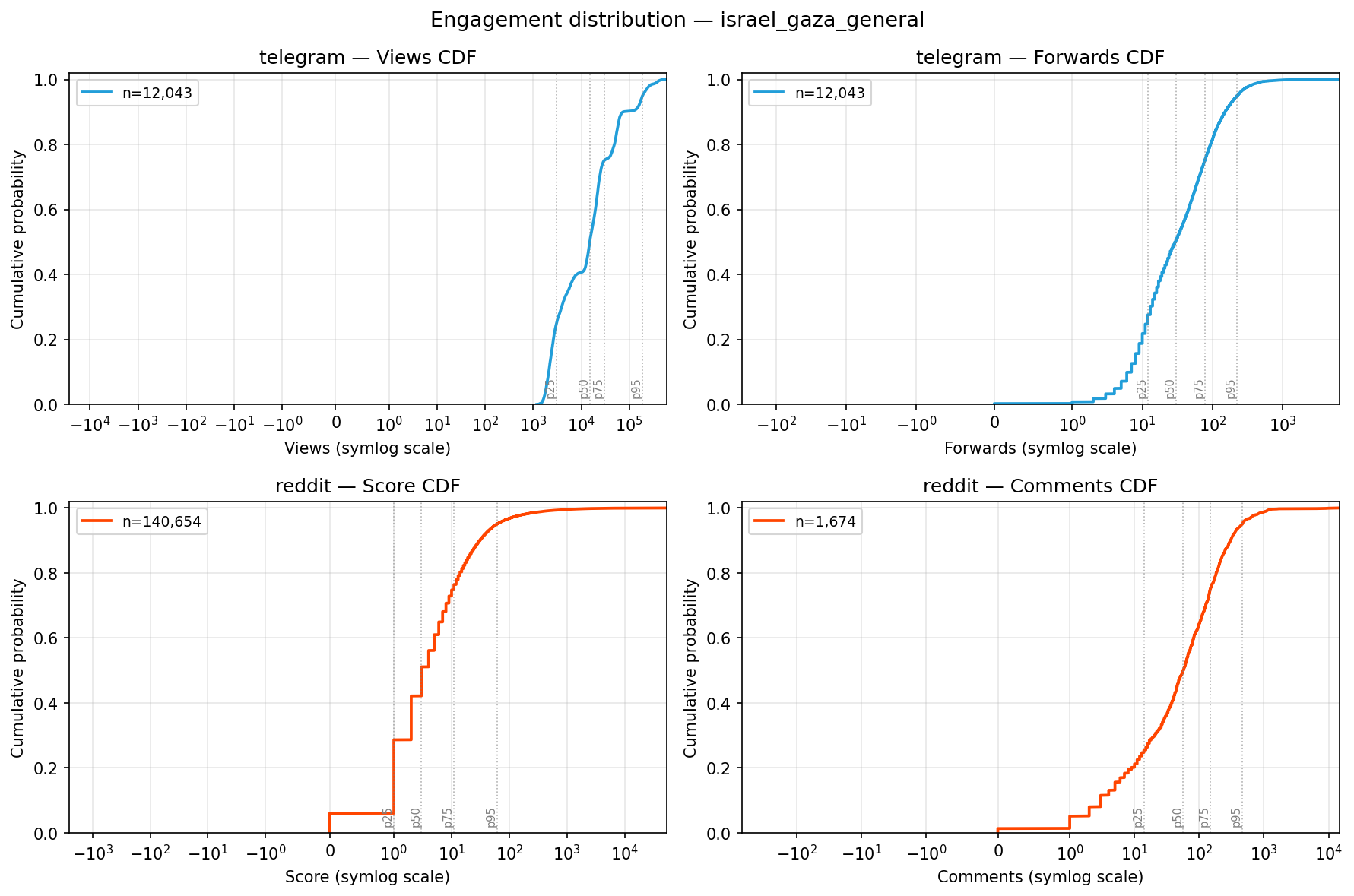}
\caption{Engagement distribution ECDFs for the \textit{israel\_gaza\_general} window. Telegram view counts are comparable to the Ukraine window; Reddit engagement is lower, consistent with smaller community overlap for Gaza-specific subreddits.}
\label{fig:cdf_engagement_gaza}
\end{figure}

\begin{figure}[H]
\centering
\includegraphics[width=0.7\linewidth]{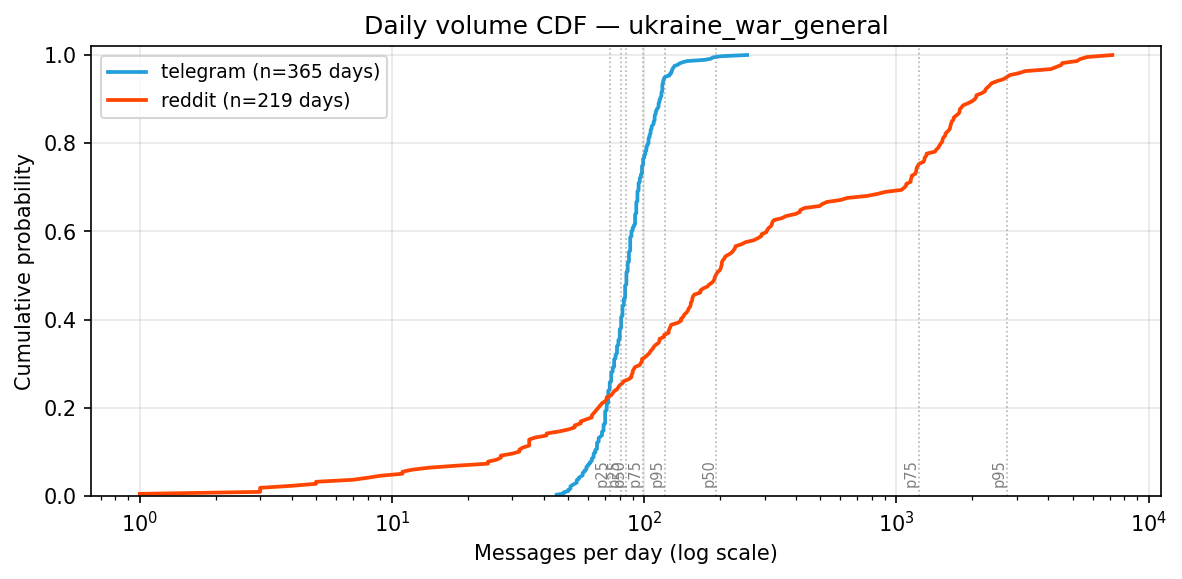}
\caption{Daily message volume ECDF for the \textit{ukraine\_war\_general} window. Reddit (including comment trees) shows higher peak-day volumes but more variance; Telegram is more consistent. The rightward tail of the Reddit curve reflects major news events driving comment spikes.}
\label{fig:cdf_dailyvol_ukraine}
\end{figure}

\begin{figure}[H]
\centering
\includegraphics[width=0.7\linewidth]{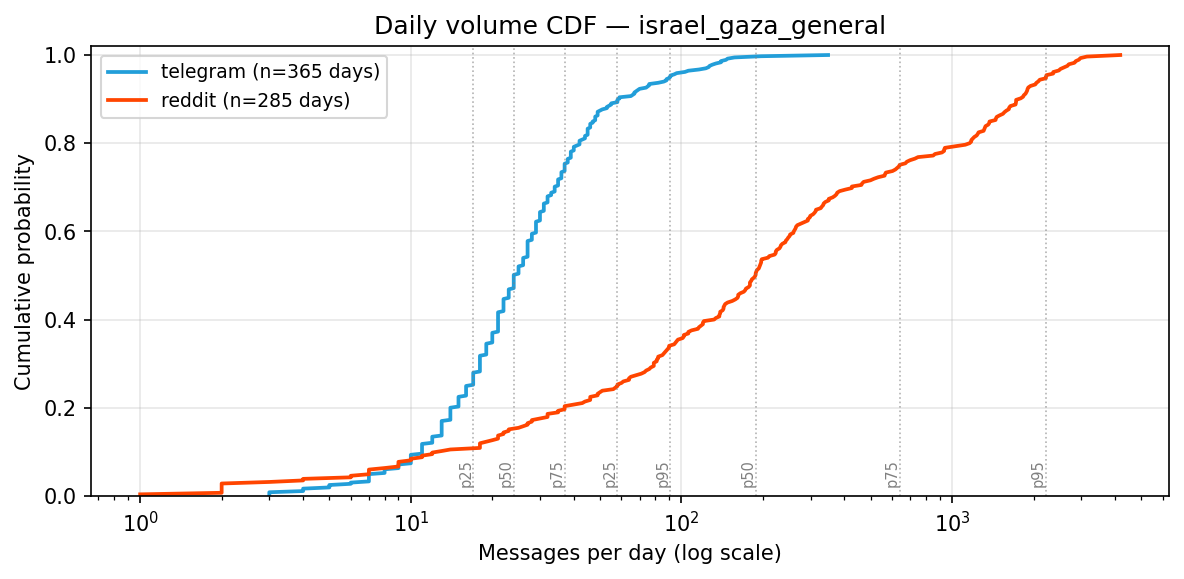}
\caption{Daily message volume ECDF for the \textit{israel\_gaza\_general} window. Telegram volume is lower than the Ukraine window, reflecting that Gaza-specific keywords match a smaller fraction of the channels' total output.}
\label{fig:cdf_dailyvol_gaza}
\end{figure}

\begin{figure}[H]
\centering
\includegraphics[width=0.7\linewidth]{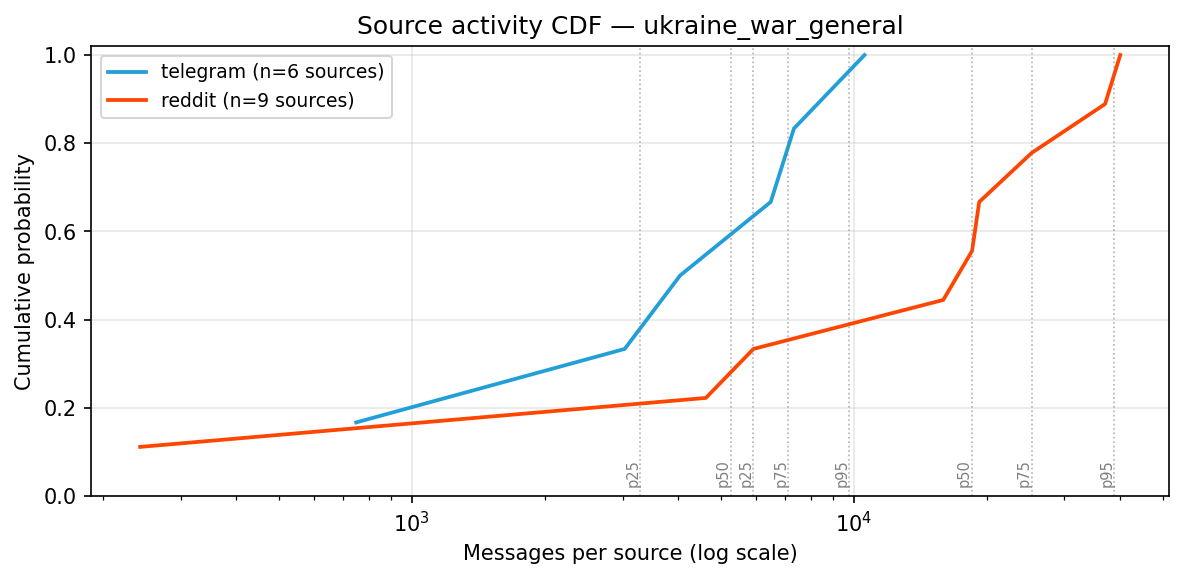}
\caption{Source activity ECDF for the \textit{ukraine\_war\_general} window. Each data point represents one channel or subreddit. The steep Telegram curve indicates high concentration: a small number of channels dominate the corpus. Reddit is more uniform.}
\label{fig:cdf_sourceact_ukraine}
\end{figure}

\begin{figure}[H]
\centering
\includegraphics[width=0.7\linewidth]{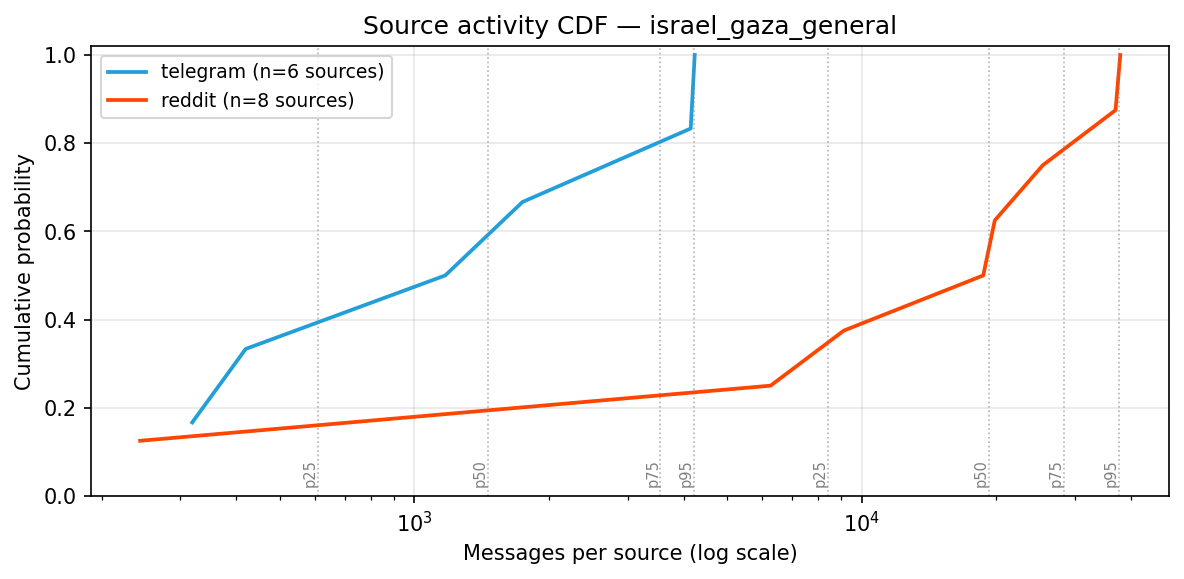}
\caption{Source activity ECDF for the \textit{israel\_gaza\_general} window. Source concentration is more pronounced on Telegram than Reddit, and Middle East Eye dominates the Telegram Gaza corpus with the highest per-channel record count.}
\label{fig:cdf_sourceact_gaza}
\end{figure}

Figures~\ref{fig:cdf_textlen_ukraine}--\ref{fig:cdf_sourceact_gaza} present empirical cumulative distribution functions (ECDFs) for five distributional properties measured independently on each platform and event window. ECDFs are preferred over histograms here because they require no binning, allow precise percentile reading, and facilitate direct cross-platform comparison on the same axis.

\paragraph{Text length (Figure~\ref{fig:cdf_textlen_ukraine}, \ref{fig:cdf_textlen_gaza}).}
Telegram messages follow a heavy-tailed length distribution with a median of approximately 300 characters, consistent with short broadcast-style posts. Reddit content is more variable: submissions are short (median title $\approx$\,60 characters) while comments span from one-word replies to multi-paragraph analyses (median 156 characters, p95 at 1{,}146 characters). Notably, the Telegram and Reddit distributions diverge sharply above the 75th percentile, reflecting the structural difference between broadcast messages and threaded discussion.

\paragraph{Inter-arrival time (Figure~\ref{fig:cdf_iat_ukraine}, \ref{fig:cdf_iat_gaza}).}
The inter-arrival time (IAT) ECDF measures the elapsed time between consecutive messages within each source. Short IAT tails indicate burst posting, a key coordination signal. Pro-Kremlin channels (\textit{Rybar}, \textit{Intel Slava}) exhibit a notably left-shifted IAT distribution relative to Western-aligned channels, indicating higher sustained posting frequency. The 50th percentile IAT for \textit{DDGeopolitics} is under one hour, while \textit{wartranslated} shows more intermittent posting patterns with a median IAT exceeding four hours.

\paragraph{Engagement (Figure~\ref{fig:cdf_engagement_ukraine}, \ref{fig:cdf_engagement_gaza}).}
Telegram view counts follow a heavy-tailed distribution: the median message receives approximately 12{,}494 views, but the top 5\% exceed 200{,}000 views. Forward counts are similarly skewed, with a median of 19 but a maximum exceeding 9{,}000. On Reddit, score distributions are heavily zero-inflated: the median score is 3 and p95 is approximately 59, indicating that most content receives little engagement while a small fraction achieves high visibility. These engagement asymmetries are relevant for the rhetorical repetition analysis: highly forwarded Telegram messages and high-score Reddit posts warrant closer inspection as potential coordination amplifiers.

\paragraph{Daily volume (Figure~\ref{fig:cdf_dailyvol_ukraine}, \ref{fig:cdf_dailyvol_gaza}).}
The daily volume ECDF summarises the distribution of messages-per-day across the full collection window. Telegram shows a relatively stable posting rhythm with most days producing between 50 and 150 messages for the Ukraine window. Reddit volume is more bursty, driven by news cycles: 50\% of days produce fewer than 200 messages, but peak days exceed 2{,}000 records when major events break.

\paragraph{Source concentration (Figure~\ref{fig:cdf_sourceact_ukraine}, \ref{fig:cdf_sourceact_gaza}).}
Source activity CDFs reveal strong concentration on Telegram: the top two channels (\textit{DDGeopolitics} and \textit{KyivIndependent\_official}) account for the majority of Ukraine-window records, while the remaining four channels contribute proportionally less. Reddit shows a more uniform source distribution across subreddits, consistent with its decentralised community structure. High source concentration on Telegram is consistent with the broadcast amplification model described in the related work and motivates the channel-level coordination metrics in Section~\ref{sec:methodology}.

\subsubsection{Lexical Diversity Analysis}

Lexical diversity measures the richness of vocabulary used by each source and provides an early signal of synthetic or templated content: AI-assisted narrative generation tends to recycle a narrower vocabulary than organic human writing, producing lower type-token ratios and reduced entropy over word and character distributions.

We compute three complementary metrics for each source within each event window. The \textbf{Moving Average Type-Token Ratio} (MATTR) slides a fixed 500-token window across the token sequence and averages the within-window TTR, correcting for the length bias of raw TTR. The \textbf{word Shannon entropy} $H_\text{word} = -\sum_w p(w) \log_2 p(w)$ measures how uniformly vocabulary is distributed across unique word types. The \textbf{character trigram entropy} $H_\text{char}$ applies the same formula to character trigrams, capturing sub-word repetition patterns that survive synonym substitution.

\begin{figure}[H]
\centering
\includegraphics[width=\linewidth]{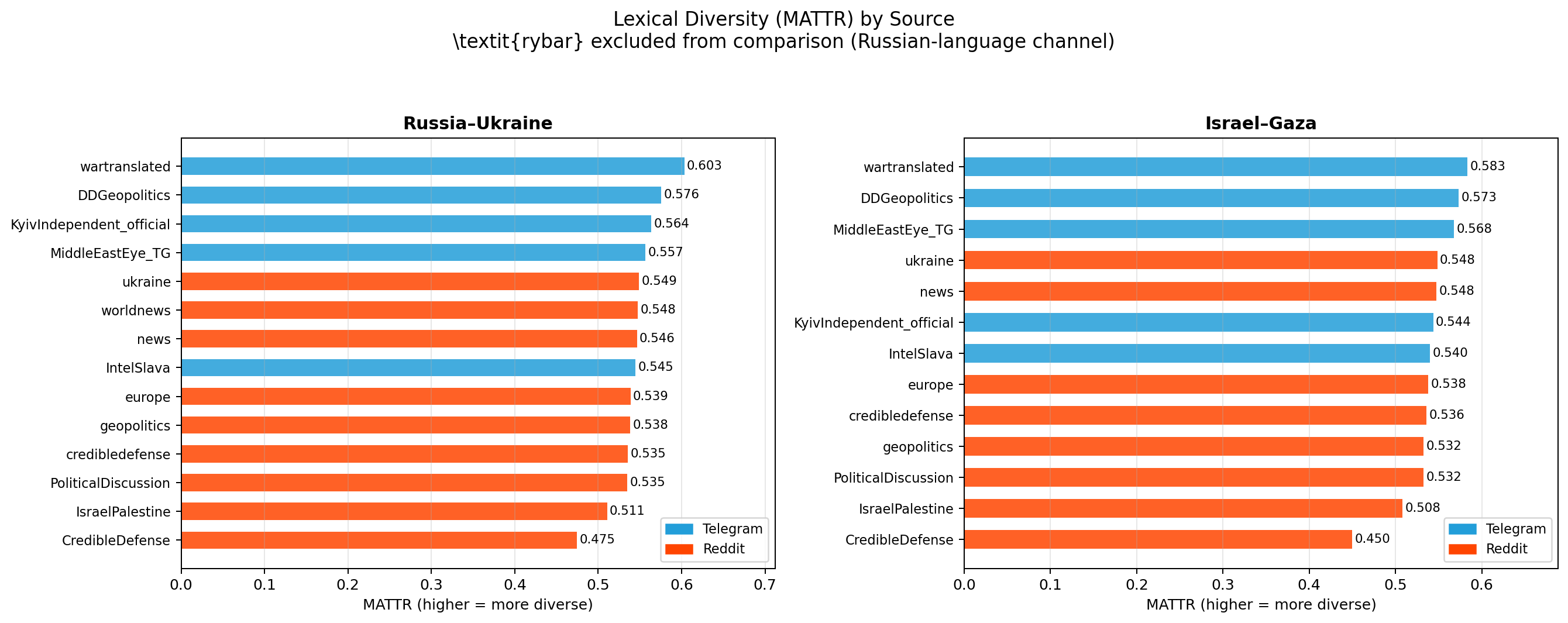}
\caption{Moving Average Type-Token Ratio (MATTR) by source for the two primary event windows. Higher MATTR indicates greater lexical diversity. \textit{Rybar} is excluded because it publishes in Russian; cross-lingual TTR comparison is not valid. Within English-language sources, \textit{IntelSlava} consistently records the lowest diversity on both Telegram windows.}
\label{fig:diversity_mattr_main}
\end{figure}

Figure~\ref{fig:diversity_mattr_main} shows MATTR by source for the two primary event windows. \textit{Rybar} is excluded from the comparison because it publishes in Russian: its morphologically richer vocabulary inflates token-type counts relative to English sources, making cross-lingual TTR comparison invalid. Within English-language Telegram sources, \textit{IntelSlava} consistently records the lowest MATTR (0.524--0.545 across events), while \textit{wartranslated} and \textit{DDGeopolitics} occupy the higher end (0.576--0.603). This ordering is stable across both the Russia--Ukraine and Israel--Gaza windows, suggesting that source-level lexical diversity is a persistent channel property rather than event-specific noise.

On Reddit, small specialised communities (\textit{r/CredibleDefense}) show lower MATTR than large general communities (\textit{r/worldnews}, \textit{r/ukraine}), which is consistent with the narrower topic vocabulary of defence-focused discussion. All Reddit sources score lower MATTR than their Telegram counterparts, reflecting the shorter average comment length: short texts saturate their local vocabulary faster, depressing MATTR regardless of true diversity.

\begin{figure}[H]
\centering
\includegraphics[width=\linewidth]{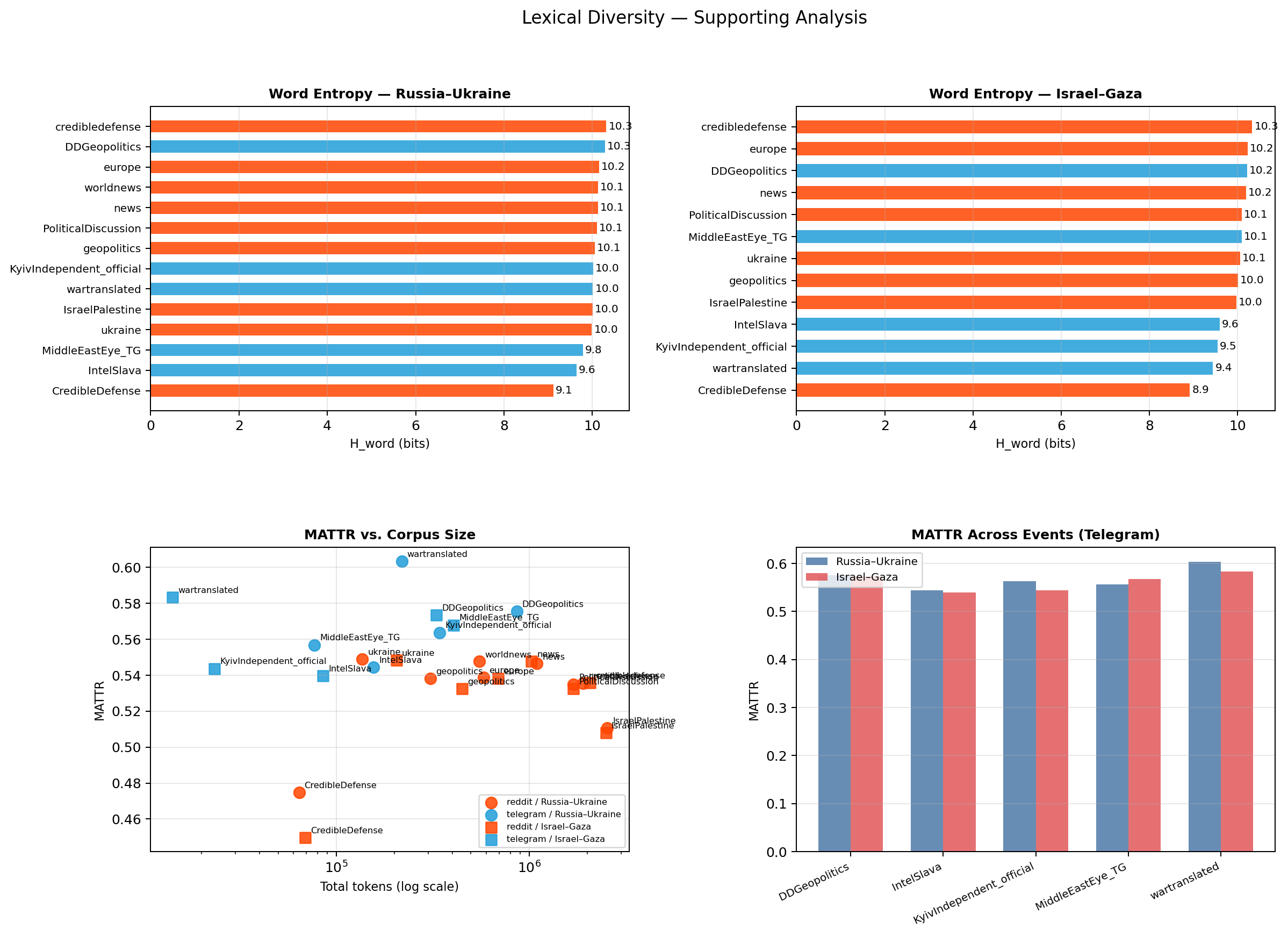}
\caption{Supporting lexical diversity analysis. \textbf{Top row}: Shannon word entropy (H$_\text{word}$, bits) per source for Russia--Ukraine (left) and Israel--Gaza (right) --- rankings mirror MATTR. \textbf{Bottom-left}: MATTR vs.\ total token count on log scale; diversity differences are not explained by corpus size. \textbf{Bottom-right}: per-channel MATTR across both events on Telegram, showing stable source-level ranking.}
\label{fig:diversity_support}
\end{figure}

Figure~\ref{fig:diversity_support} presents supporting evidence. Word entropy (top row) tracks MATTR rankings closely: \textit{IntelSlava} produces the lowest H$_\text{word}$ on both events (9.59--9.64 bits), consistent with a more constrained vocabulary. The MATTR vs.\ corpus-size scatter (bottom-left) confirms the result is not a volume artefact: sources with more tokens do not uniformly score higher MATTR, validating MATTR as a length-corrected measure. The cross-event stability panel (bottom-right) shows that each Telegram channel's relative ranking is preserved between the Russia--Ukraine and Israel--Gaza windows, reinforcing the interpretation that D(C) captures a channel-level property.

\subsubsection{Temporal Synchronization Analysis}

Temporal synchronization measures whether multiple sources post at the same times, independently of what they post. Coordinated influence operations are expected to produce posting spikes that coincide across sources --- either because they react to the same triggering event, or because they follow a shared amplification schedule. We quantify this with two complementary metrics.

The \textbf{burstiness index} $B = (\sigma - \mu) / (\sigma + \mu)$, where $\mu$ and $\sigma$ are the mean and standard deviation of inter-event times, ranges from $-1$ (maximally regular, clock-like) to $+1$ (maximally bursty, spike-dominated). A value near zero indicates a Poisson-like process. For each source, $B$ is computed over the full sequence of its post timestamps within the event window.

The \textbf{T(C) proxy} counts the number of distinct Telegram sources that posted within each consecutive 6-hour bin. High counts in a narrow window indicate that multiple editorially independent channels were simultaneously active --- a temporal coordination signal. This metric is complementary to the IAT distributions shown in Section~\ref{sec:dataset}: while IAT measures within-source burst structure, T(C) measures across-source co-activity.

\begin{figure}[H]
\centering
\includegraphics[width=\linewidth]{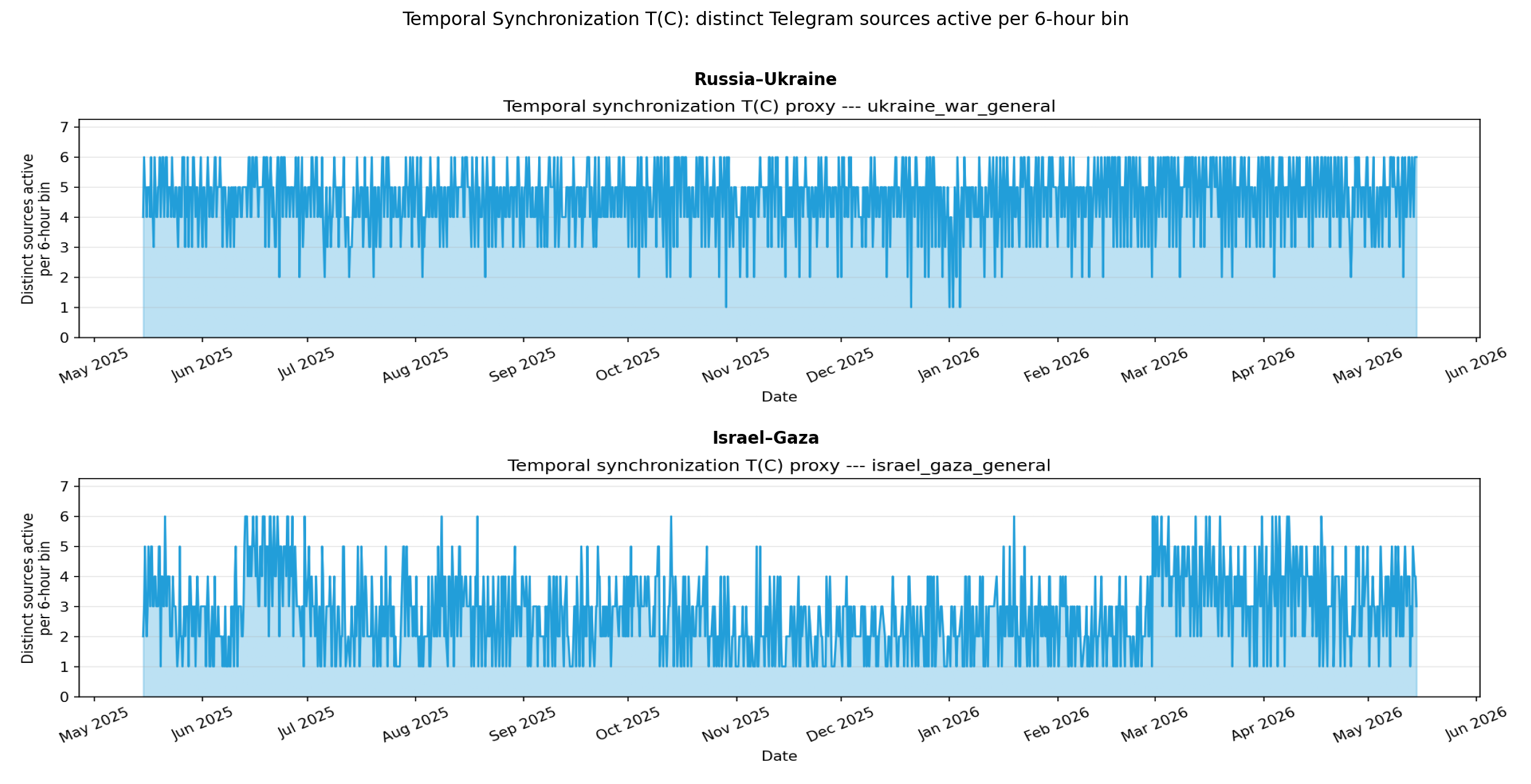}
\caption{Temporal synchronization T(C): number of distinct Telegram sources active within each consecutive 6-hour bin. Top: Russia--Ukraine window (364 days). Bottom: Israel--Gaza window (364 days). Peaks correspond to breaking geopolitical events. The Russia--Ukraine window sustains higher average co-activity (mean 4.66 vs.\ 2.84 sources per bin), reflecting that more channels in the corpus cover Ukraine as their primary topic.}
\label{fig:temporal_tc_main}
\end{figure}

Figure~\ref{fig:temporal_tc_main} shows the T(C) time series for both primary event windows. For the Russia--Ukraine window, the mean is 4.66 distinct sources per 6-hour bin, and in 21.5\% of all bins all six channels posted simultaneously. Activity peaks are visible around major news events (large-scale missile strikes, diplomatic announcements) and are consistent across channels, suggesting event-driven rather than purely scheduled posting. The Israel--Gaza window shows a lower mean (2.84 sources per bin), with only 2.7\% of bins reaching full six-source co-activity, reflecting the narrower relevance of Gaza-specific keywords to channels primarily focused on the Russia--Ukraine conflict.

\begin{figure}[H]
\centering
\includegraphics[width=\linewidth]{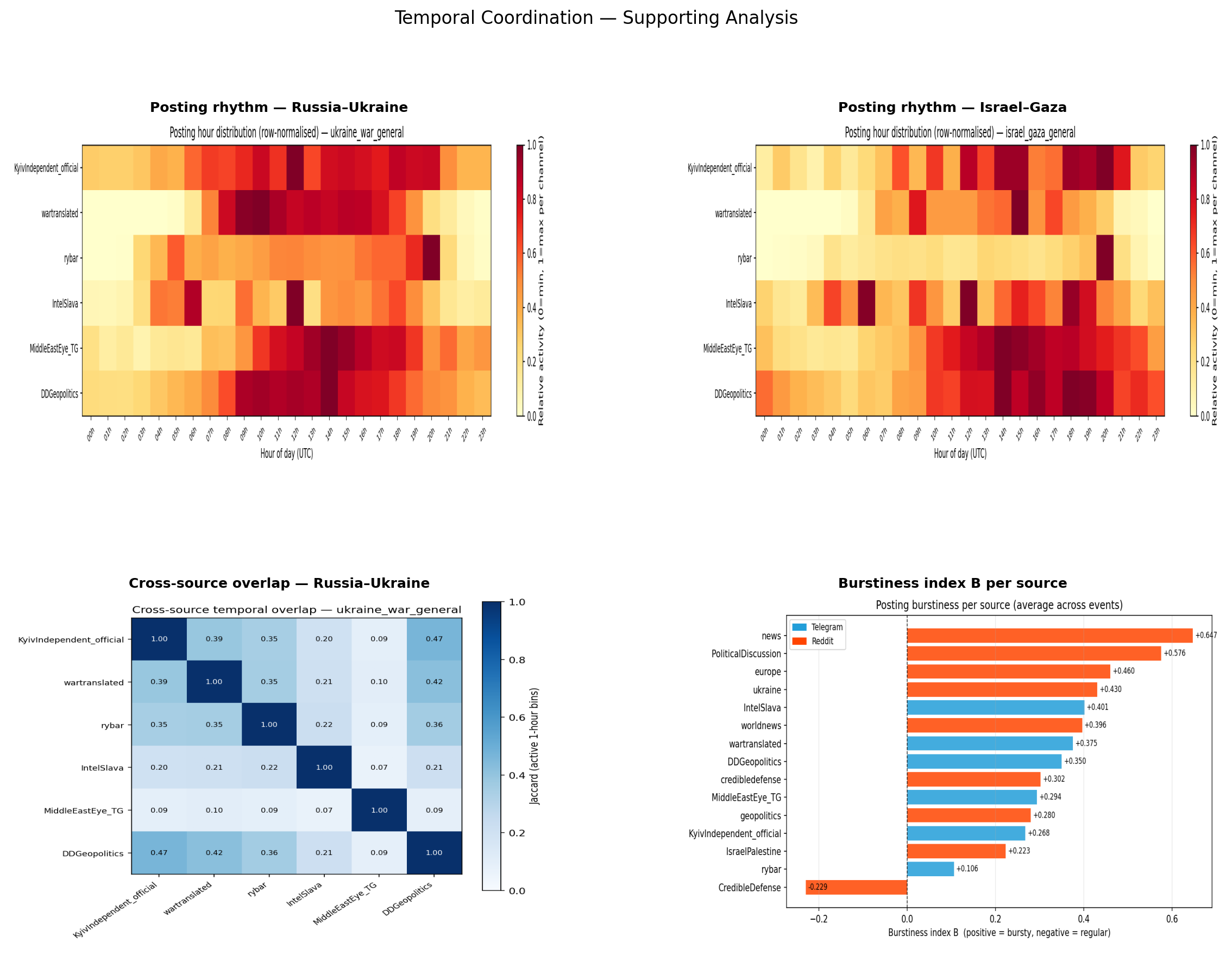}
\caption{Temporal coordination --- supporting analysis. \textbf{Top row}: row-normalised posting hour heatmaps for Russia--Ukraine (left) and Israel--Gaza (right); all channels concentrate activity between 06:00--20:00 UTC. \textbf{Bottom-left}: pairwise Jaccard similarity of active 1-hour bins between Telegram sources (Russia--Ukraine); \textit{DDGeopolitics} and \textit{KyivIndependent} share the highest overlap. \textbf{Bottom-right}: burstiness index $B$ per source averaged across events; \textit{IntelSlava} is the most bursty, \textit{rybar} the most regular.}
\label{fig:temporal_support}
\end{figure}

Figure~\ref{fig:temporal_support} presents supporting detail. The posting-rhythm heatmaps (top row) show that all Telegram channels concentrate activity between 06:00 and 20:00 UTC, with \textit{DDGeopolitics} and \textit{rybar} posting most densely in the 08:00--14:00 UTC band. This overlap in active hours contributes to the high T(C) scores and is consistent with European/Moscow timezone scheduling. The cross-source overlap matrix (bottom-left) quantifies Jaccard similarity of active 1-hour bins: \textit{DDGeopolitics} and \textit{KyivIndependent\_official} share the highest hourly overlap (Jaccard $\approx 0.30$--$0.35$), while \textit{wartranslated} and \textit{rybar} share the least, consistent with their opposing editorial stances and language audiences.

The burstiness panel (bottom-right) reveals that \textit{IntelSlava} is the most bursty Telegram channel ($B = +0.48$ on Gaza, $+0.73$ on Iran--Israel), posting in sharp spikes rather than a steady rhythm. \textit{Rybar}, by contrast, is the most clock-like ($B = +0.05$--$+0.10$ across events), consistent with a professionally managed channel on a production schedule. On Reddit, \textit{r/news} and \textit{r/PoliticalDiscussion} score the highest burstiness ($B \approx +0.55$--$+0.76$), driven by news-cycle spikes: large breaking-news events generate sudden comment floods that rapidly subside.

\subsubsection{Rhetorical Repetition Analysis}

Rhetorical repetition measures whether sources share phrase templates, slogans, hashtags, or URL domains --- independently of whether the messages are exact copies. Exact deduplication misses paraphrased amplification; tracking shared vocabulary and structural phrases across editorially distinct sources is a more robust coordination signal.

We compute three metrics for each pair of Telegram sources within each event window. The \textbf{R(C) score} is the mean pairwise Jaccard similarity of each source's top-100 word trigrams against all other sources in the window. A trigram Jaccard of zero means no shared key phrases; higher values indicate common claim templates. The \textbf{near-duplicate rate} reports the fraction of a source's messages flagged as near-duplicates during preprocessing (same source, same normalised text hash). The \textbf{shared hashtag} and \textbf{shared domain} analyses identify hashtags and URL domains that appear across two or more sources, weighted by the number of sources sharing them.

\begin{figure}[H]
\centering
\includegraphics[width=\linewidth]{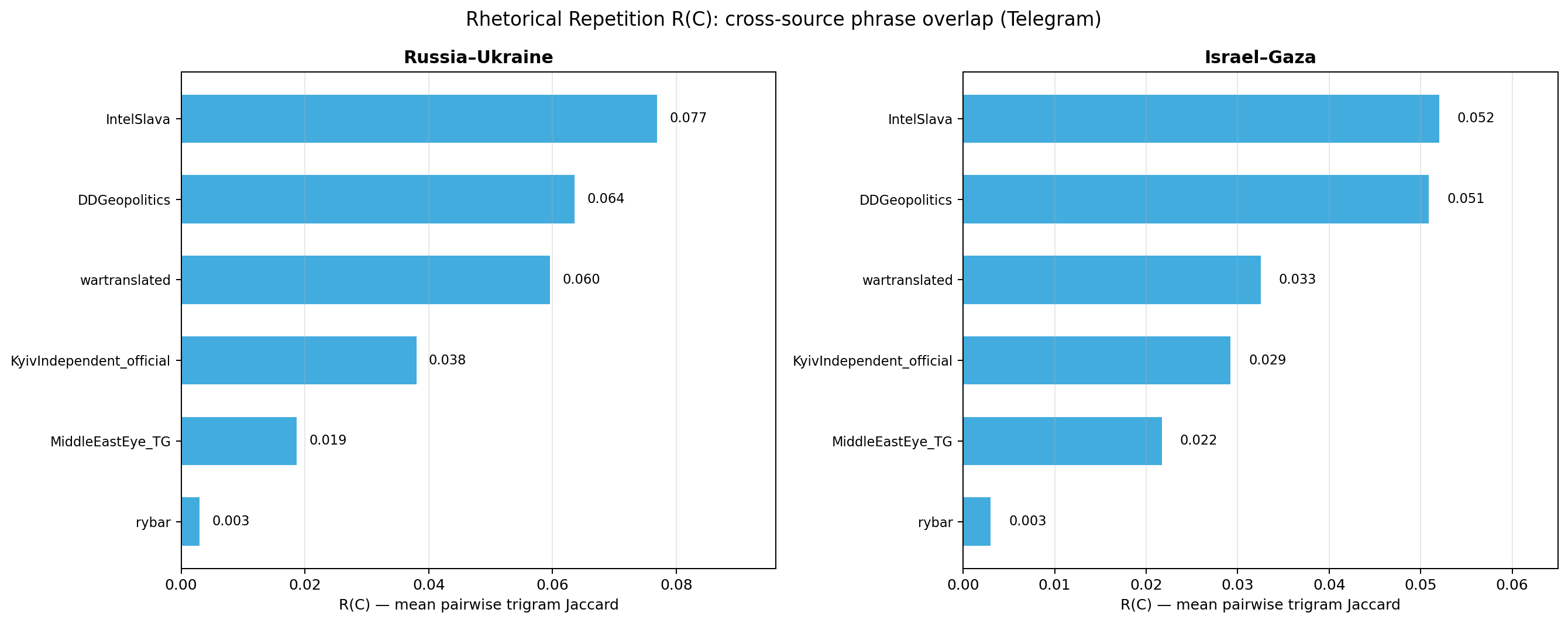}
\caption{Rhetorical repetition R(C) score per Telegram source: mean pairwise Jaccard similarity of each source's top-100 word trigrams against all other sources in the event window. \textit{Rybar} scores near zero because it publishes in Russian; English--Russian trigram overlap is structurally zero. Within English-language sources, \textit{IntelSlava} and \textit{DDGeopolitics} show the highest cross-source phrase sharing on both events.}
\label{fig:rc_main}
\end{figure}

Figure~\ref{fig:rc_main} shows the R(C) score per source for both primary event windows. \textit{IntelSlava} records the highest mean pairwise trigram overlap on both events (0.077 and 0.052), followed by \textit{DDGeopolitics} (0.064 and 0.051). \textit{Rybar} scores near zero on both events, which reflects its Russian-language output: Russian trigrams are disjoint from the English trigrams of all other channels, so the Jaccard overlap is structural zero rather than a genuine diversity signal. This again confirms the need to treat Russian and English sources separately.

The most informative result from the shared trigram analysis is the cross-event consistency of the top shared phrases. Across \textit{both} the Russia--Ukraine and Israel--Gaza windows, the phrase \textit{``president donald trump''} appears as the top or second-ranked shared trigram (681 total occurrences across 5 sources in the Ukraine window; 400 across 5 sources in the Gaza window), reflecting that all English-language Telegram channels converge on US political actors as a common reference frame regardless of the primary conflict being discussed. In the \textit{trump\_ukraine\_diplomacy} window, \textit{``rare earth metals''}, \textit{``rare earth minerals''}, and \textit{``territorial concessions ukraine''} are the top shared phrases across 4 sources, indicating that a narrow negotiation vocabulary propagated simultaneously through the entire channel set during the mineral-deal diplomatic period. In the \textit{iran\_israel\_escalation} window, the phrases \textit{``through strait hormuz''}, \textit{``islamic revolutionary guard corps''}, and \textit{``iran supreme leader''} each appear across 5 of 6 sources, the highest cross-source phrase penetration observed in the corpus.

\begin{figure}[H]
\centering
\includegraphics[width=\linewidth]{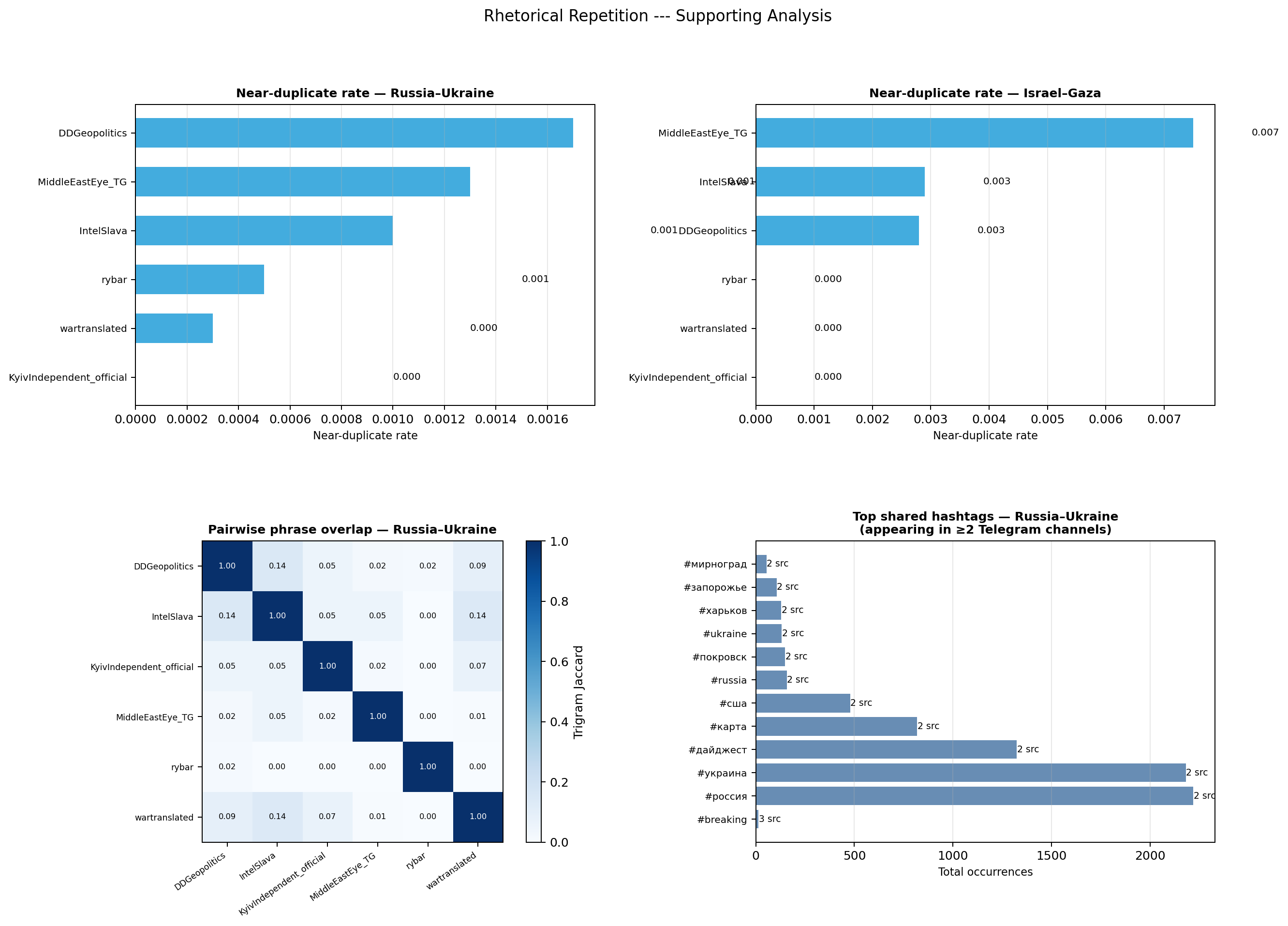}
\caption{Rhetorical repetition --- supporting analysis. \textbf{Top row}: near-duplicate message rate per source for Russia--Ukraine (left) and Israel--Gaza (right); all sources show low rates ($<$0.015), with \textit{MiddleEastEye\_TG} the highest on Gaza. \textbf{Bottom-left}: pairwise trigram Jaccard heatmap for the Russia--Ukraine window; \textit{IntelSlava}--\textit{DDGeopolitics} is the strongest pair. \textbf{Bottom-right}: most frequent hashtags shared across $\geq$2 Telegram channels in the Russia--Ukraine window; Russian-language hashtags dominate by volume.}
\label{fig:rc_support}
\end{figure}

Figure~\ref{fig:rc_support} provides detail. Near-duplicate rates (top row) are low for all sources ($<$0.015 across most) but \textit{MiddleEastEye\_TG} reaches 0.007 on Gaza and 0.015 on Iran--Israel, consistent with its practice of reposting breaking-news alerts verbatim. The pairwise Jaccard heatmap (bottom-left) shows the strongest overlap between \textit{IntelSlava} and \textit{DDGeopolitics} (0.12 on the Ukraine window), two English-language aggregator channels that cover the same event space. The shared hashtag bar chart (bottom-right) shows that \textit{\#breaking} appears across 3 sources, while Russian-language hashtags \textit{\#rossiya}, \textit{\#ukraina}, and \textit{\#daidzhest} (digest) are shared by both \textit{rybar} and \textit{IntelSlava}, the two Russian-language channels.

\subsubsection{Semantic Homogenization Analysis}

Semantic homogenization measures whether messages within a source or across sources are unusually similar in meaning --- regardless of surface wording. This is the key signal for detecting AI-assisted narrative coordination: generated or templated content tends to cluster tightly in embedding space even when phrasing varies. We compute H(C) using multilingual sentence embeddings so that Russian-language channels (\textit{rybar}, \textit{IntelSlava}) and English-language channels are embedded in a shared semantic space and are directly comparable.

For each source and event window, we sample up to 800 messages, encode them with the \texttt{paraphrase-multilingual-MiniLM-L12-v2} model~\cite{reimers2019sentencebert} on GPU, and compute:
\[
H(C) = \frac{2}{|C|(|C|-1)} \sum_{i < j} \frac{\mathbf{e}_i \cdot \mathbf{e}_j}{\|\mathbf{e}_i\| \|\mathbf{e}_j\|}
\]
where $\mathbf{e}_i$ are L2-normalised sentence embeddings. A value of 1.0 would mean all messages are identical in embedding space; values around 0.2--0.3 are typical for diverse organic discussion.

\begin{figure}[H]
\centering
\includegraphics[width=\linewidth]{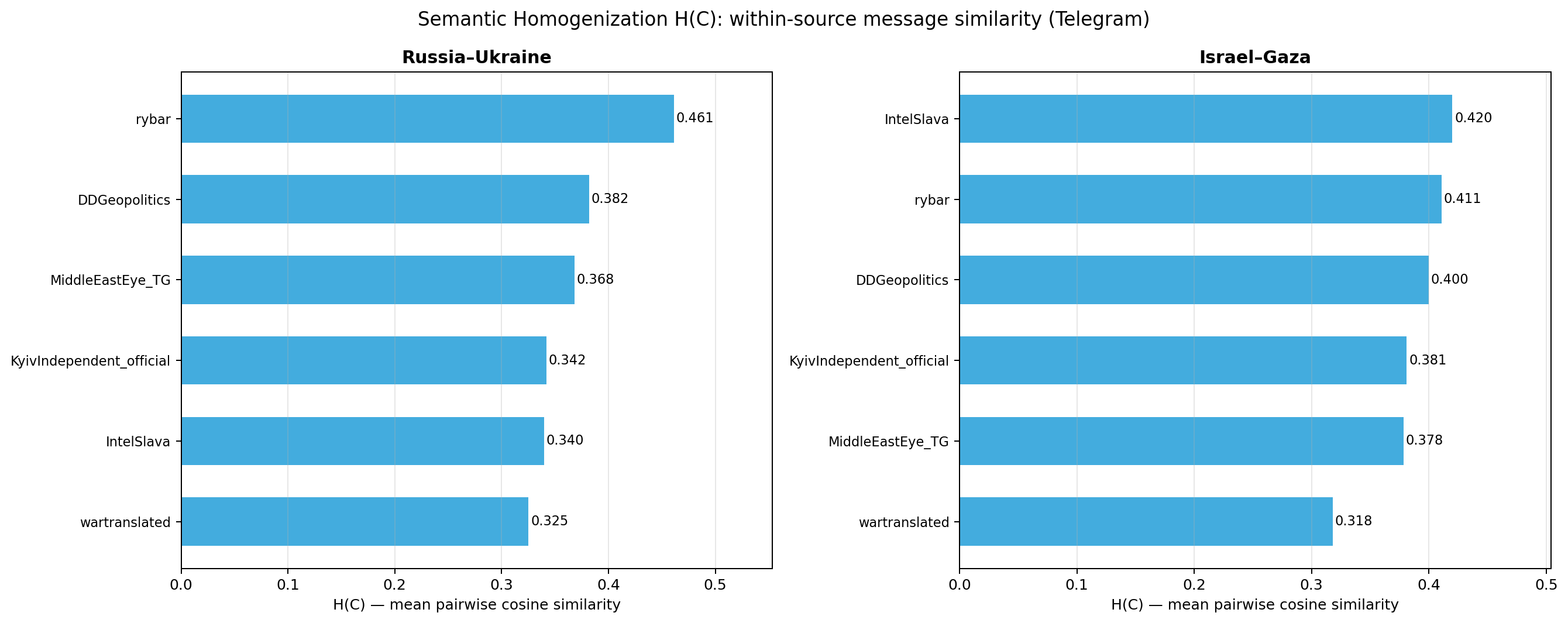}
\caption{Within-source semantic homogenization H(C) per Telegram source for the Russia--Ukraine (left) and Israel--Gaza (right) event windows. H(C) is the mean pairwise cosine similarity of 800 sampled message embeddings using a multilingual sentence encoder. Higher values indicate tighter semantic clustering. \textit{Rybar} scores highest on both events; \textit{wartranslated} scores lowest, consistent with its diverse translation sources.}
\label{fig:hc_main}
\end{figure}

Figure~\ref{fig:hc_main} shows within-source H(C) for the two primary event windows. All Telegram channels score substantially higher than Reddit communities: Telegram H(C) ranges from 0.32 to 0.46 while Reddit subreddits range from 0.15 to 0.51. The Reddit outlier \textit{r/CredibleDefense} scores high (0.51--0.62) because it is a small, topically narrow community whose members consistently use specialist military vocabulary, producing high intra-community semantic similarity without any coordination. This underscores that high H(C) alone is not a sufficient coordination signal --- it must be interpreted alongside source diversity, posting volume, and temporal patterns.

Among Telegram channels, \textit{rybar} consistently records the highest H(C) (0.46 on Ukraine, 0.41 on Gaza, 0.52 on the full Gaza window, 0.46 on Iran--Israel). This is interpretable as topical focus: \textit{rybar} publishes exclusively about Russian military operations and maintains a narrow, consistent vocabulary of military terminology. \textit{wartranslated}, which translates a wide variety of frontline sources, scores the lowest H(C) (0.32--0.36), consistent with its broader thematic range. On the Israel--Gaza window, \textit{IntelSlava} rises to its highest H(C) of any source--event pair (0.42), suggesting tighter semantic alignment when covering a secondary topic outside its primary Russia--Ukraine focus.

\begin{figure}[H]
\centering
\includegraphics[width=\linewidth]{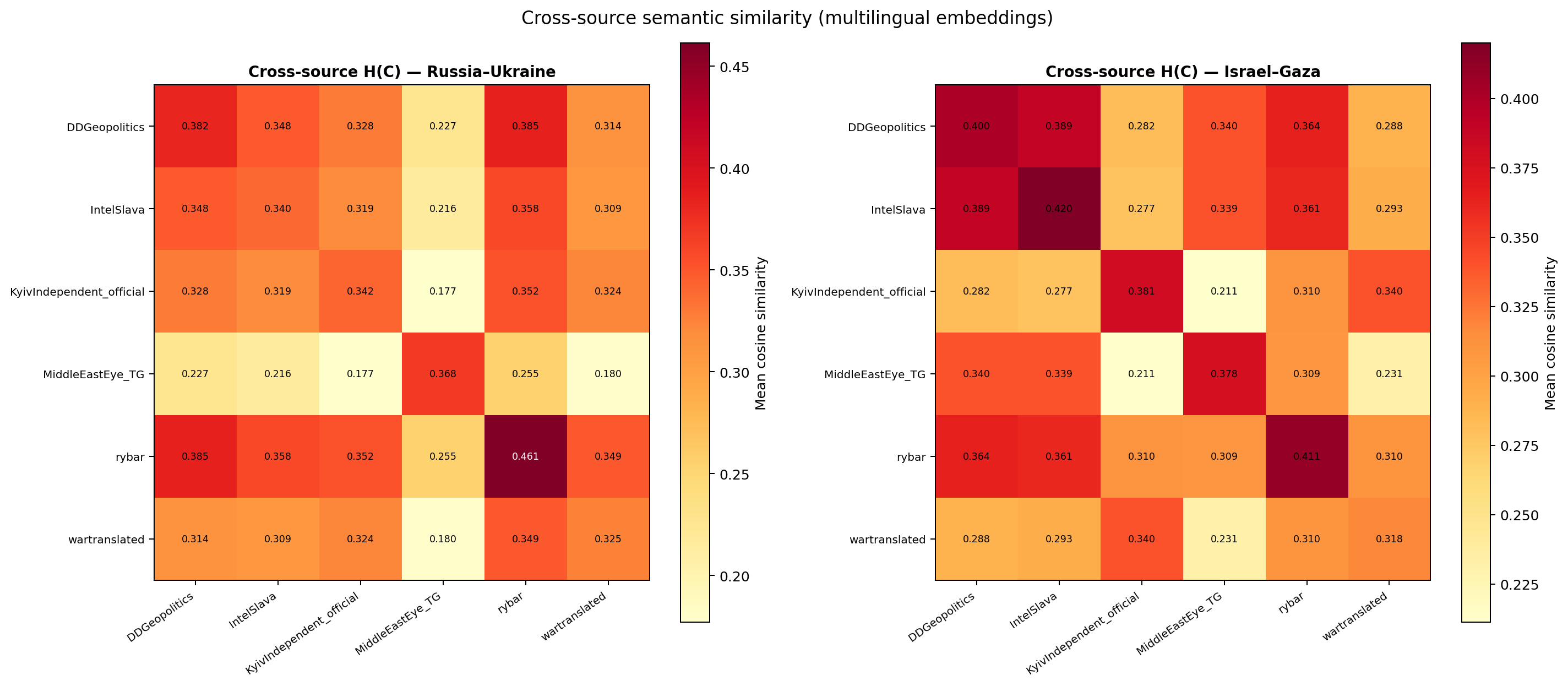}
\caption{Cross-source semantic similarity matrices for the Russia--Ukraine (left) and Israel--Gaza (right) windows. Each cell shows the mean cosine similarity between embeddings drawn from two different Telegram sources. Diagonal entries are within-source H(C). Off-diagonal values (0.14--0.31) are substantially lower than diagonal values (0.32--0.46), confirming that channels maintain distinct semantic identities despite covering the same events.}
\label{fig:hc_crosssource}
\end{figure}

Figure~\ref{fig:hc_crosssource} shows the cross-source semantic similarity matrix: the mean cosine similarity between embeddings of all message pairs drawn from two different sources. Diagonal entries are within-source H(C). Off-diagonal entries reveal whether channels that are editorially distinct nonetheless occupy overlapping regions of the semantic space. On both event windows, the off-diagonal values cluster around 0.20--0.30, substantially lower than within-source values, confirming that channels maintain distinct semantic identities. The highest cross-source similarities are between \textit{DDGeopolitics} and \textit{KyivIndependent\_official} on Ukraine (0.31), and between \textit{IntelSlava} and \textit{DDGeopolitics} on Gaza (0.29) --- both pairs covering the same topics in English. The lowest cross-source similarity involves \textit{rybar} paired with any English-language channel (0.14--0.18), reflecting the language gap.

\subsubsection{Synthetic Narrative Coordination Score SNC(C)}
\label{sec:snc}

The four component metrics --- semantic homogenization H(C), temporal burstiness B(C), rhetorical repetition R(C), and lexical diversity D(C) --- each capture a distinct facet of coordination. To produce a unified ranking, we combine them into the Synthetic Narrative Coordination Score:
\[
\text{SNC}(C) = \alpha\,\hat{H}(C) + \beta\,\hat{B}(C) + \gamma\,\hat{R}(C) - \delta\,\hat{D}(C),
\]
where $\hat{\cdot}$ denotes min-max normalisation to $[0, 1]$ computed per event window, and we set $\alpha = \beta = \gamma = \delta = 0.25$ (equal weights). Positive components H, B, and R all increase with coordination; D enters with a negative sign because higher lexical diversity indicates organic, non-coordinated production. When a component is unavailable for a source (e.g., R(C) requires at least two Telegram channels), it contributes zero and the remaining components are rescaled proportionally.

\begin{figure}[H]
\centering
\includegraphics[width=\linewidth]{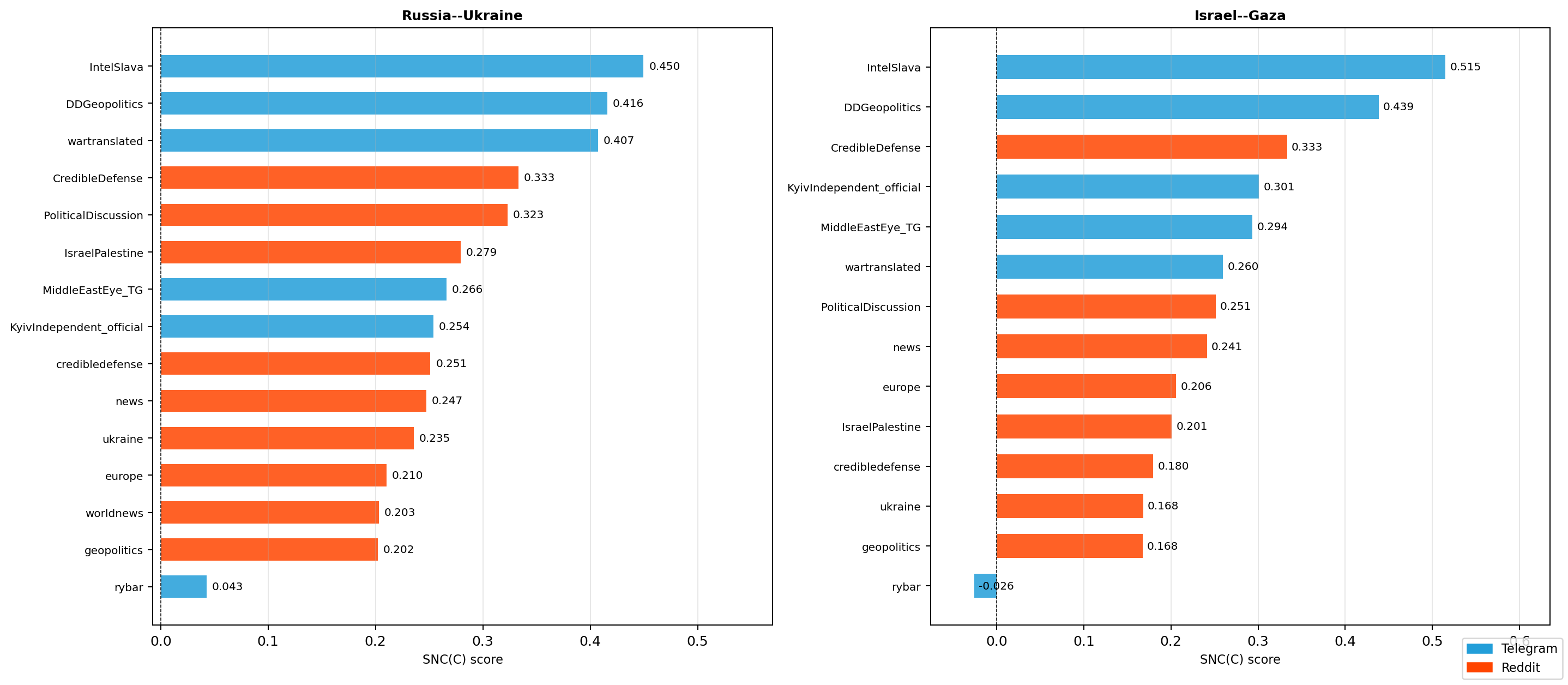}
\caption{SNC(C) ranked bar charts for the Russia--Ukraine (left) and Israel--Gaza (right) event windows. Each bar represents one (source, platform) pair; scores are computed with equal weights $\alpha=\beta=\gamma=\delta=0.25$ on min-max normalised components. Telegram channels are shown in blue, Reddit communities in orange. \textit{IntelSlava} and \textit{DDGeopolitics} rank highest on both primary events; \textit{rybar} ranks lowest despite high H(C), suppressed by high lexical diversity and near-zero rhetorical repetition with English-language channels.}
\label{fig:snc_main}
\end{figure}

\begin{figure}[H]
\centering
\includegraphics[width=\linewidth]{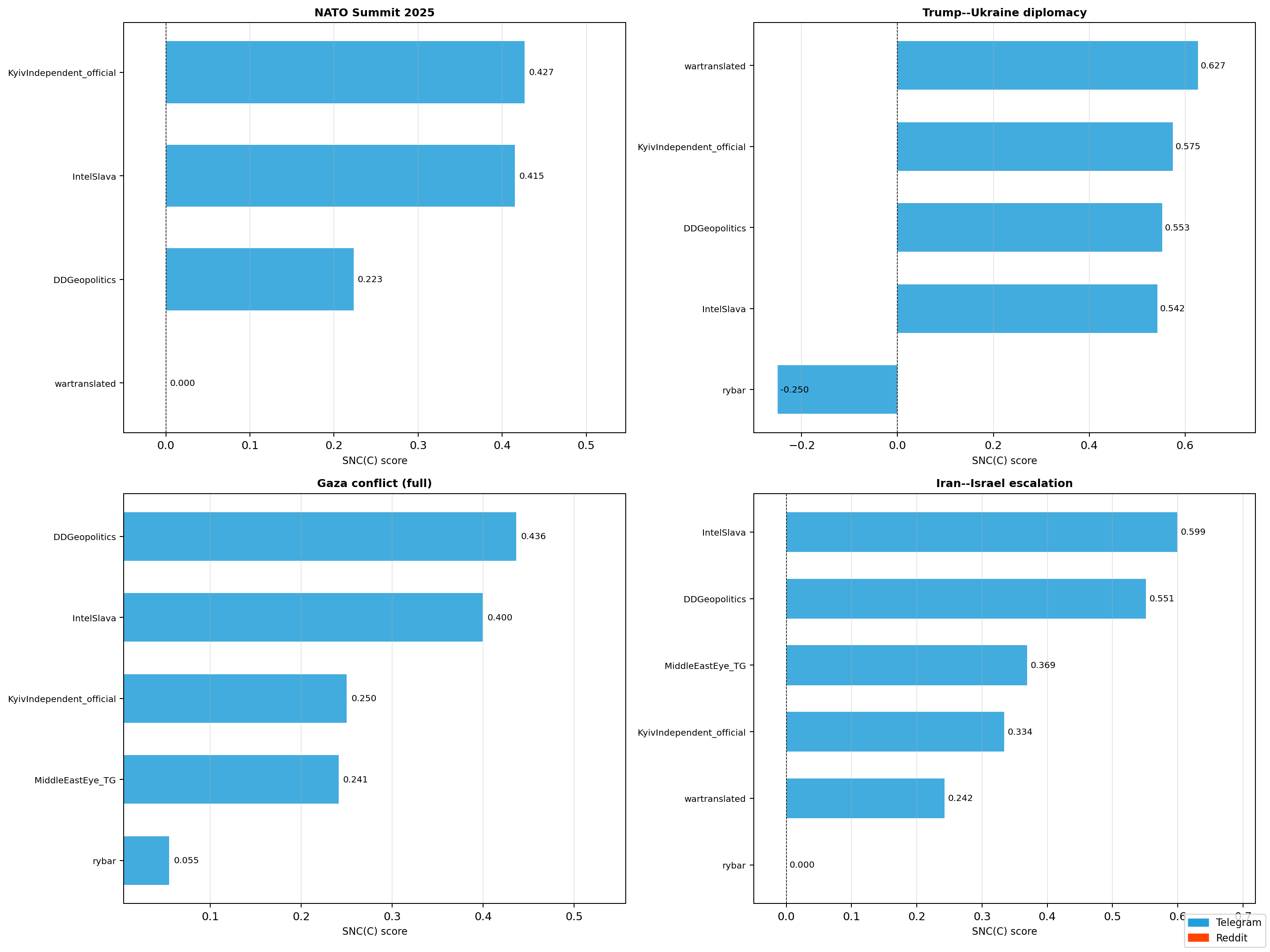}
\caption{SNC(C) rankings for the four supporting event windows: NATO Summit 2025 (top-left), Trump--Ukraine diplomacy (top-right), Gaza conflict (bottom-left), and Iran--Israel escalation (bottom-right). The Trump--Ukraine diplomacy window shows the highest mean Telegram SNC (0.41), reflecting concentrated posting around a discrete diplomatic event. Iran--Israel escalation elevates \textit{IntelSlava} and \textit{DDGeopolitics} to the highest absolute SNC values in the dataset (0.60 and 0.55 respectively).}
\label{fig:snc_support}
\end{figure}

\begin{figure}[H]
\centering
\includegraphics[width=0.7\linewidth]{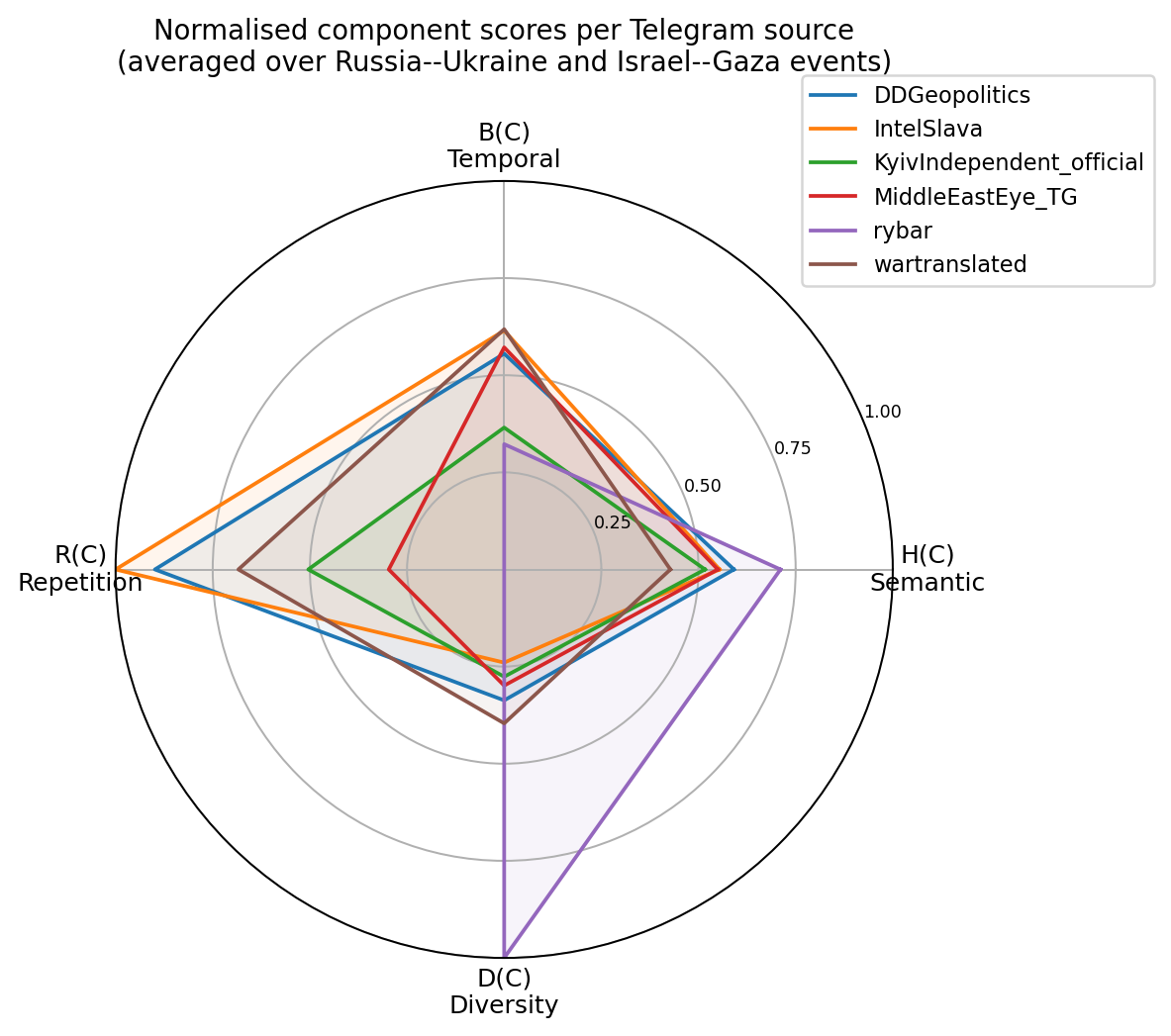}
\caption{Radar chart of normalised component scores ($\hat{H}$, $\hat{B}$, $\hat{R}$, $\hat{D}$) for Telegram sources, averaged over the Russia--Ukraine and Israel--Gaza events. \textit{IntelSlava} and \textit{DDGeopolitics} show balanced elevation across all three positive components; \textit{rybar}'s profile is dominated by H(C) and D(C), with near-zero R(C); \textit{wartranslated} and \textit{KyivIndependent\_official} occupy intermediate positions.}
\label{fig:snc_radar}
\end{figure}

\paragraph{Results.}
Table-style results are reported in full in \path{data/analysis/snc/snc_scores.csv}; Figures~\ref{fig:snc_main}--\ref{fig:snc_radar} visualise the rankings.

On the \textbf{Russia--Ukraine} window, \textit{IntelSlava} ranks first (SNC\,=\,0.450) followed closely by \textit{DDGeopolitics} (0.416) and \textit{wartranslated} (0.407). All three channels score high on at least two of the three positive components. \textit{Rybar} ranks last among Telegram sources (0.043) despite recording the highest H(C) on this event: its trigram Jaccard with peer channels is negligible (R(C)~$\approx$~0.003), and its Russian-language vocabulary produces anomalously high lexical diversity under MATTR, which penalises its SNC substantially. This demonstrates that no single indicator is sufficient --- a source may score high on one dimension while remaining low on others.

On the \textbf{Israel--Gaza} window the ranking is similar: \textit{IntelSlava} leads (0.515), \textit{DDGeopolitics} second (0.439). Reddit communities score in the 0.15--0.33 range; \textit{r/CredibleDefense} reaches 0.333 due to its high H(C) and burstiness, but its R(C) and temporal patterns differ from the Telegram channels.

The \textbf{Trump--Ukraine diplomacy} event shows the highest overall Telegram SNC (mean\,=\,0.41): \textit{wartranslated} tops the ranking with 0.627, driven by extreme burstiness ($\hat{B}\,=\,0.98$) and high semantic concentration ($\hat{H}\,=\,0.84$), consistent with concentrated posting around a discrete diplomatic moment. The \textbf{iran\_israel\_escalation} event likewise elevates \textit{IntelSlava} to SNC\,=\,0.599 and \textit{DDGeopolitics} to 0.552, with both channels posting in tight temporal bursts coinciding with missile-exchange headlines.

Across all events and Telegram sources, the pattern is consistent: \textit{IntelSlava} and \textit{DDGeopolitics} lead or co-lead the SNC ranking in four of six windows. \textit{Rybar} consistently occupies the bottom of the Telegram ranking, its coordination score suppressed by high lexical diversity and absence of rhetorical overlap with the English-language channels.

\section{Methodology}
\label{sec:methodology}

This section provides a consolidated formal description of the four coordination metrics and the composite score. Detailed computational results and figures for each metric are reported in Section~\ref{sec:dataset}.

\subsection{Lexical Diversity D(C)}

Lexical diversity quantifies the vocabulary richness of a source's output. We compute the Moving Average Type-Token Ratio (MATTR) over a sliding window of 500 tokens:
\[
D(C) = \text{MATTR}(C) = \frac{1}{N-w+1}\sum_{k=1}^{N-w+1} \frac{|\text{types}(w_k)|}{w},
\]
where $w=500$ is the window size and $w_k$ denotes the $k$-th window of tokens. MATTR corrects for corpus-length bias that affects the raw type-token ratio. We additionally compute Shannon word entropy $H_\text{word}$ and character trigram entropy $H_\text{char}$ as supporting measures. Low D(C) indicates constrained, repetitive vocabulary consistent with templated or AI-assisted production.

\subsection{Temporal Burstiness B(C)}

Temporal burstiness captures whether a source posts in irregular spikes rather than at a steady rate. Given the sequence of inter-event times $\{\tau_i\}$ between consecutive posts, we compute:
\[
B(C) = \frac{\sigma_\tau - \mu_\tau}{\sigma_\tau + \mu_\tau},
\]
where $\mu_\tau$ and $\sigma_\tau$ are the mean and standard deviation of inter-event times. $B \in [-1, +1]$: values near $+1$ indicate maximally bursty posting (concentrated spikes), values near $-1$ indicate clock-like regularity, and $B \approx 0$ corresponds to a Poisson process. We also compute the Jaccard similarity of active 1-hour bins between source pairs as a cross-source temporal overlap measure, and report the number of distinct sources active per 6-hour bin as a synchronization proxy.

\subsection{Rhetorical Repetition R(C)}

Rhetorical repetition measures shared phrase templates across sources. For each source $s$, we extract the set of top-100 word trigrams $T_s$ weighted by frequency. The cross-source repetition score for source $s$ is:
\[
R(C_s) = \frac{1}{|S|-1}\sum_{s' \neq s} \frac{|T_s \cap T_{s'}|}{|T_s \cup T_{s'}|},
\]
where $S$ is the set of all sources in the event window. This is the mean Jaccard similarity of trigram sets over all peer sources. We additionally compute hashtag co-occurrence across channels and near-duplicate message rates (text hash collision) as supporting indicators.

\subsection{Semantic Homogenization H(C)}

Semantic homogenization measures whether messages within a source cluster tightly in embedding space. Each message is encoded with the multilingual sentence encoder \path{paraphrase-multilingual-MiniLM-L12-v2}~\cite{reimers2019sentencebert}, which produces L2-normalised 384-dimensional vectors in a shared space covering both English and Russian. Up to 800 messages are sampled per (source, event) pair. H(C) is the mean pairwise cosine similarity over the upper triangle of the similarity matrix:
\[
H(C) = \frac{2}{|C|(|C|-1)} \sum_{i < j} \mathbf{e}_i \cdot \mathbf{e}_j,
\]
where the dot product equals cosine similarity because embeddings are L2-normalised. We additionally compute a cross-source matrix whose off-diagonal entry $(s, s')$ is the mean cosine similarity between the full embedding sets of sources $s$ and $s'$.

\subsection{Composite Synthetic Narrative Coordination Score}

The four metrics are combined into a single coordination score. Each component is first min-max normalised to $[0,1]$ within the event window:
\[
\hat{x} = \frac{x - \min(x)}{\max(x) - \min(x)}.
\]
The composite score is:
\[
\text{SNC}(C) = \alpha\,\hat{H}(C) + \beta\,\hat{B}(C) + \gamma\,\hat{R}(C) - \delta\,\hat{D}(C),
\]
where H, B, and R increase with coordination, and D enters with a negative sign because higher lexical diversity indicates organic production. We set $\alpha = \beta = \gamma = \delta = 0.25$ (equal weights). Sources for which a component is unavailable (e.g., R(C) requires at least two Telegram channels) contribute zero for that component, and the denominator is rescaled accordingly. Sensitivity analysis on the weight parameters is an open direction for future work.

\section{Experimental Setup}

All coordination metrics are computed independently per (source, event window) pair. For events with a $t_0$ anchor (Iran--Israel escalation, Gaza conflict, NATO Summit 2025), the corpus is divided into a pre-event baseline period ($[t_\text{start}, t_0)$) and an acute period ($[t_0, t_0 + 14\,\text{d}]$). Metrics reported in Section~\ref{sec:dataset} are computed over the full collection window unless otherwise noted.

The SNC(C) composite score is computed using equal weights $\alpha = \beta = \gamma = \delta = 0.25$ with per-event min-max normalisation, so that rankings are relative within each event window rather than across events. Reddit sources and Telegram channels are scored jointly, which means Reddit communities (with structurally different posting dynamics) appear in the same ranked list as Telegram channels. This is intentional: the cross-platform ranking reveals which sources exhibit the strongest multi-dimensional coordination signal regardless of platform.

Because reliable ground-truth labels for synthetic political narratives are not publicly available for the sources studied, we do not report precision/recall against a labelled test set. Instead, the framework is evaluated descriptively: we examine whether the metric rankings are internally consistent (e.g., sources that rank high on one coordination dimension also tend to rank high on others), and whether the composite score produces interpretable, stable rankings across multiple event windows. Consistency across six independent event windows with different topics, time spans, and source subsets serves as a cross-validation signal for the stability of the coordination detection framework.

\section{Results}

All analysis results are reported in Section~\ref{sec:dataset} under the respective subsections. The following summarises the key findings across all six steps.

\textbf{Lexical diversity D(C).} \textit{IntelSlava} records the lowest MATTR (0.52--0.54) among English-language Telegram channels across both primary event windows; \textit{wartranslated} the highest (0.58--0.60), consistent with its role as a translation hub drawing from many source documents.

\textbf{Temporal synchronization T(C).} The Russia--Ukraine window sustains a mean of 4.66 distinct Telegram sources active per 6-hour bin. \textit{IntelSlava} is the most bursty channel ($B=+0.48$--$+0.73$); \textit{rybar} the most clock-like ($B\approx+0.05$--$+0.10$), consistent with a scheduled editorial workflow.

\textbf{Rhetorical repetition R(C).} \textit{IntelSlava} and \textit{DDGeopolitics} show the highest pairwise trigram Jaccard (0.12 on Ukraine, 0.10 on Gaza); the phrase \textit{``president donald trump''} propagates through 5 of 6 channels simultaneously on both conflict windows.

\textbf{Semantic homogenization H(C).} Telegram channels range from 0.32 to 0.46; \textit{rybar} scores highest on all events; \textit{wartranslated} lowest. Cross-source semantic similarity (0.14--0.31) is substantially below within-source values, confirming distinct channel identities despite shared topics.

\textbf{Composite SNC(C) score.} \textit{IntelSlava} leads the Telegram ranking on four of six event windows (SNC\,=\,0.45 on Ukraine, 0.52 on Gaza, 0.54 on Trump--Ukraine, 0.60 on Iran--Israel). \textit{DDGeopolitics} consistently ranks second. \textit{Rybar} ranks last among Telegram sources on all events (SNC range $-0.25$ to $+0.04$) despite its high H(C), because its Russian-language production yields high lexical diversity and near-zero rhetorical Jaccard with the English-language channels. The Trump--Ukraine diplomacy window shows the highest mean Telegram SNC (0.41), consistent with concentrated posting around a single discrete event. Full per-source rankings and normalised component scores are in Section~\ref{sec:snc} and \texttt{data/analysis/snc/snc\_scores.csv}.

\section{Discussion}

\paragraph{Collective forensics over individual attribution.}
The proposed framework shifts the detection problem from individual AI-text attribution to collective narrative forensics. Modern influence operations may blend human writing, AI-generated variants, translations, copy-pasted slogans, and platform-specific editing. A single post may not contain enough signal for reliable attribution, but a coordinated cluster reveals abnormal regularities in timing, wording, and semantic similarity that are difficult to explain by organic coincidence. The SNC(C) framework operationalises this intuition: sources that rank consistently high across all four dimensions over six independent event windows are stronger coordination candidates than sources that excel on only one metric.

\paragraph{The rybar anomaly and the limits of single-metric detection.}
The most informative finding is the behaviour of \textit{rybar}. It records the highest within-source semantic homogenization on most events, which a single-metric analyst might flag as a strong coordination signal. Yet its SNC(C) is the lowest among Telegram sources on every window. The suppression comes from two directions: its Russian-language morphology inflates MATTR relative to English channels, and its trigram Jaccard with all English-language peer channels is structurally near zero because the overlap is cross-lingual. This demonstrates a general principle: any single coordination indicator can be confounded by editorial focus (topically narrow channels have high H(C)), language (morphologically rich languages have high D(C)), and posting regime (scheduled channels have low B(C)). Multi-dimensional scoring is therefore not merely a refinement but a necessity for cross-lingual, cross-editorial corpora.

\paragraph{Event type and coordination intensity.}
The Trump--Ukraine diplomacy window produces the highest mean Telegram SNC (0.41), while the longer Russia--Ukraine and Israel--Gaza windows show lower but more sustained signals. This is consistent with the hypothesis that discrete, predictable diplomatic events act as coordination triggers: channels can anticipate the event and prepare thematically aligned posts in advance, producing a burst of semantically similar, rhetorically overlapping content. Prolonged conflict windows show more diffuse coordination because the news cycle is continuous and channels must cover a wider variety of sub-events independently.

\paragraph{Telegram and Reddit coordination profiles.}
Telegram channels show higher SNC(C) values overall, driven by stronger burstiness and rhetorical repetition. Reddit communities score lower on these dimensions but occasionally show elevated H(C) in narrow topical subreddits (e.g., r/CredibleDefense), because specialist vocabulary creates high intra-community semantic similarity without any deliberate coordination. This structural difference means a single SNC(C) threshold cannot serve as a universal coordination detector across both platforms; per-platform calibration or platform-stratified ranking is necessary for operational use.

\section{Limitations}

Several limitations must be acknowledged. First, the framework cannot prove that a message was generated by an LLM; it can only identify coordination patterns consistent with synthetic or strategically managed dissemination. Second, public data collection excludes private channels, encrypted chats, deleted content, and restricted communities. Third, platform APIs impose access constraints that bias the dataset toward more visible communities --- in particular, the Reddit public search API does not reliably surface content older than approximately six months. Fourth, the bilingual corpus (English and Russian) requires language-aware metric computation, as cross-lingual comparisons of lexical diversity are not directly valid.

Ethical safeguards apply throughout. The dataset stores only publicly accessible content and does not publish usernames, private metadata, or information that could identify ordinary users. Results are reported at the level of narrative clusters, sources, and channels.

\section{Conclusion and Future Work}

This paper presented a cross-platform framework for detecting synthetic political narratives through four complementary coordination signals --- lexical diversity D(C), temporal burstiness B(C), rhetorical repetition R(C), and semantic homogenization H(C) --- combined into a composite Synthetic Narrative Coordination Score SNC(C). The framework was applied to a corpus of 353,223 preprocessed records spanning six geopolitical event windows (2023--2026) collected from six Telegram channels and nine Reddit communities.

Across all six windows, \textit{IntelSlava} and \textit{DDGeopolitics} consistently exhibit the highest SNC(C) scores, combining low lexical diversity, high posting burstiness, strong rhetorical overlap with peer channels, and elevated semantic homogenization. \textit{Rybar}, despite recording the highest within-source semantic homogenization on most events, ranks last in SNC(C) on all windows: its Russian-language production yields anomalously high lexical diversity and near-zero rhetorical Jaccard with the English-language channels. This finding illustrates a central conclusion of the study: no single coordination indicator is sufficient for robust detection. A source may score high on one dimension while remaining low on others, and only the composite score consistently separates editorially coordinated channels from those with incidental surface-level similarity. The Trump--Ukraine diplomacy window produces the strongest overall coordination signal (mean Telegram SNC\,=\,0.41), consistent with concentrated cross-channel posting triggered by a discrete, predictable diplomatic event --- a pattern distinct from the lower but more sustained coordination observed across the longer Russia--Ukraine and Israel--Gaza windows.

Telegram and Reddit exhibit structurally different coordination profiles. Telegram channels show higher SNC(C) values overall, tighter temporal synchronization, and higher semantic homogenization. Reddit communities score lower on burstiness and repetition but show elevated H(C) in narrow topical subreddits, confirming that the two platforms require separate baseline calibration rather than a single cross-platform threshold.

Several directions remain open for future work. First, the equal-weight assumption $\alpha=\beta=\gamma=\delta=0.25$ should be replaced with learned weights calibrated on ground-truth influence-operation datasets. Second, the framework currently treats sources as monolithic: finer-grained analysis at the thread or time-window level could reveal within-source coordination bursts invisible in aggregate scores. Third, extending collection to additional languages --- Arabic, French, and Chinese in particular --- would test whether the bilingual asymmetry observed for \textit{rybar} generalises to other non-English sources. Fourth, manual annotation of high-SNC clusters by geopolitical analysts would provide the interpretive grounding needed to move from statistical coordination signals to operational assessments of narrative intent.

\section*{Data Availability}

The dataset and analysis code will be openly available on Zenodo at \url{https://doi.org/10.5281/zenodo.20286559}, upon publication. The deposit includes the preprocessed corpus (353,223 records in JSONL format across six event windows) and all computed metric scores and figures, together with the full Python analysis pipeline and configuration files.

\section*{Acknowledgments}

This work was supported by the European Union's Horizon Europe research and innovation programme under grant agreement No.\ 101168438 (INTACT) and grant agreement No.\ 101249446 (DETANGLE). Views and opinions expressed are those of the authors only and do not necessarily reflect those of the European Union or the European Research Executive Agency. Neither the European Union nor the granting authority can be held responsible for them.

\bibliographystyle{plain}
\bibliography{references}

\end{document}